\documentclass{osa-article}

\journal{osajournal}

\articletype{Research Article}

\usepackage{textcomp}
\usepackage{physics}
\newcommand\unitv[1]{\hat{\vb{#1}}}

\usepackage[symbol]{footmisc}

\hypersetup{
     colorlinks   = true,
     linkcolor    = blue,
     citecolor    = blue,
     urlcolor = blue
}

\definecolor{col1}{rgb}{0,0,0}

\newcommand{\sri}[1]{{\color{col1} #1}}
\newcommand{\red}[1]{{\color{col1} #1}}

\begin{document}

\title{Inverse-designed photon extractors for optically addressable defect qubits}

\author{Srivatsa Chakravarthi,\authormark{1,5,*} Pengning Chao,\authormark{2,6,*} Christian Pederson, \authormark{3} Sean Molesky,\authormark{2}, Andrew Ivanov,\authormark{3}, Karine Hestroffer,\authormark{4} Fariba Hatami,\authormark{4} Alejandro W. Rodriguez,\authormark{2} and Kai-Mei C. Fu\authormark{1,3}}

\address{\authormark{1}Department of Electrical and Computer Engineering, University of Washington, Seattle, WA, USA\\
\authormark{2}Department of Electrical Engineering, Princeton University, Princeton, NJ, USA\\
\authormark{3}Department of Physics, University of Washington, Seattle, WA, USA\\
\authormark{4}Department of Physics, Humboldt-Universitat zu Berlin, Newtonstrasse, Berlin, Germany\\
\authormark{5}srivatsa@uw.edu\\
\authormark{6}pengning@princeton.edu}

\footnotetext[1]{These authors contributed equally to this paper.}



\begin{abstract}
Solid-state defect qubit systems with spin-photon interfaces show great promise for quantum information and metrology applications. Photon collection efficiency, however, presents a major challenge for defect qubits in high refractive index host materials. Inverse-design optimization of photonic devices enables unprecedented flexibility in tailoring critical parameters of a spin-photon interface including spectral response, photon polarization and collection mode. Further, the design process can incorporate additional constraints, such as fabrication tolerance and material processing limitations. Here we design and demonstrate a compact hybrid gallium phosphide on diamond inverse-design planar dielectric structure coupled to single near-surface nitrogen-vacancy centers formed by implantation and annealing. We observe up to a 14-fold broadband enhancement in photon extraction efficiency, in close agreement with simulation. We expect that such inverse-designed devices will enable realization of scalable arrays of single-photon emitters, rapid characterization of new quantum emitters, efficient sensing and heralded entanglement schemes.
\end{abstract}

\section{Introduction}
Optically addressable defect qubits in materials like diamond~\cite{jelezko_single_2006, bradac_quantum_2019} and silicon carbide~\cite{koehl_room_2011} 
are promising for the realization of a wide range of quantum technologies including single-photon generation, quantum metrology, and quantum information protocols~\cite{benjamin_brokered_2006, raussendorf_one-way_2001}. In all of these applications, photon collection efficiency is a key figure of merit. The photoluminescence (PL) detection efficiency is intrinsically limited by the non-directional emission of PL and total internal reflection between the high-index host material and its low-index environment. To enable high PL collection efficiency, photonic structures such as solid-immersion lenses~\cite{jamali_microscopic_2014}, waveguides~\cite{momenzadeh_nanoengineered_2015, radulaski_scalable_2017} and microcavities~\cite{li_coherent_2015, santori_nanophotonics_2010, gould_efficient_2016, schmidgall_frequency_2018} have been utilized. Nanophotonic devices are particularly attractive for their small footprint and potential for scalable integration. However, optimizing nanophotonic structures for efficient defect integration is often nontrivial and bespoke due to the unique constraints imposed by the targeted application for a specific defect system. For high-sensitivity quantum metrology with nitrogen vacancy (NV) centers in diamond, photonic structures should be optimized for efficient broadband extraction of PL from NVs located a few nm from the sensing surface. In contrast, for quantum information applications, photonic coupling to the sharp zero-phonon line (ZPL) of deep NV centers (>~100\,nm from the surface) and operation under the high-cooperativity (Purcell-enhanced) regime is preferable. 

Here we present a flexible inverse-design optimization framework that can reconcile a wide range of design constraints and generate planar dielectric gallium phosphide(GaP)-on-diamond photonic structures. \red{We selected GaP due to its high refractive index ($n=3.31$) and low loss at the NV emission energy.} \red{The structural patterning is constrained to the GaP layer in order to minimize perturbations of the defect environment.} For a dipole located 100\,nm from the surface and oriented perpendicular to the NV axis, an optimized 1.5\,\textmu m~$\times$~1.5\,\textmu m device is calculated to provide a \red{15.7}-fold PL enhancement of the free-space PL collection. \sri{The 100\,nm dipole depth corresponds to the surface-defect separation at which NV centers synthesised by implantation and annealing have been shown to exhibit high optical coherence\cite{chu_coherent_2014}.} Orders of magnitude larger enhancement factors are found for \sri{devices designed for} defects positioned closer to the surface. \red{Recent work\cite{Molesky20_limits, molesky_hierarchical_2020} on fundamental bounds to related design goals such as maximizing scattering cross-section or near field radiation extraction suggest that inverse-designed devices can have performance approaching the absolute theoretical limit.} 

Experimentally, we fabricate the optimized GaP-on-diamond photon extractors and observe efficient PL collection from implanted single NV centers. These versatile devices exhibit a broadband PL enhancement for wavelengths in the measured range of 575\,nm to 750\,nm with  up to a 14-fold zero-phonon-line enhancement for single NV \sri{centers}. Extensive optical characterization of the NV centers performed both before and after fabrication provides insight into changes in the local NV environment. Post-fabrication sample treatment substantially improves device NV optical stability and points to important fabrication/design considerations for future defect qubit-photonics integration.    

\begin{figure*}[!htb]
\centering
\includegraphics[width=\textwidth]{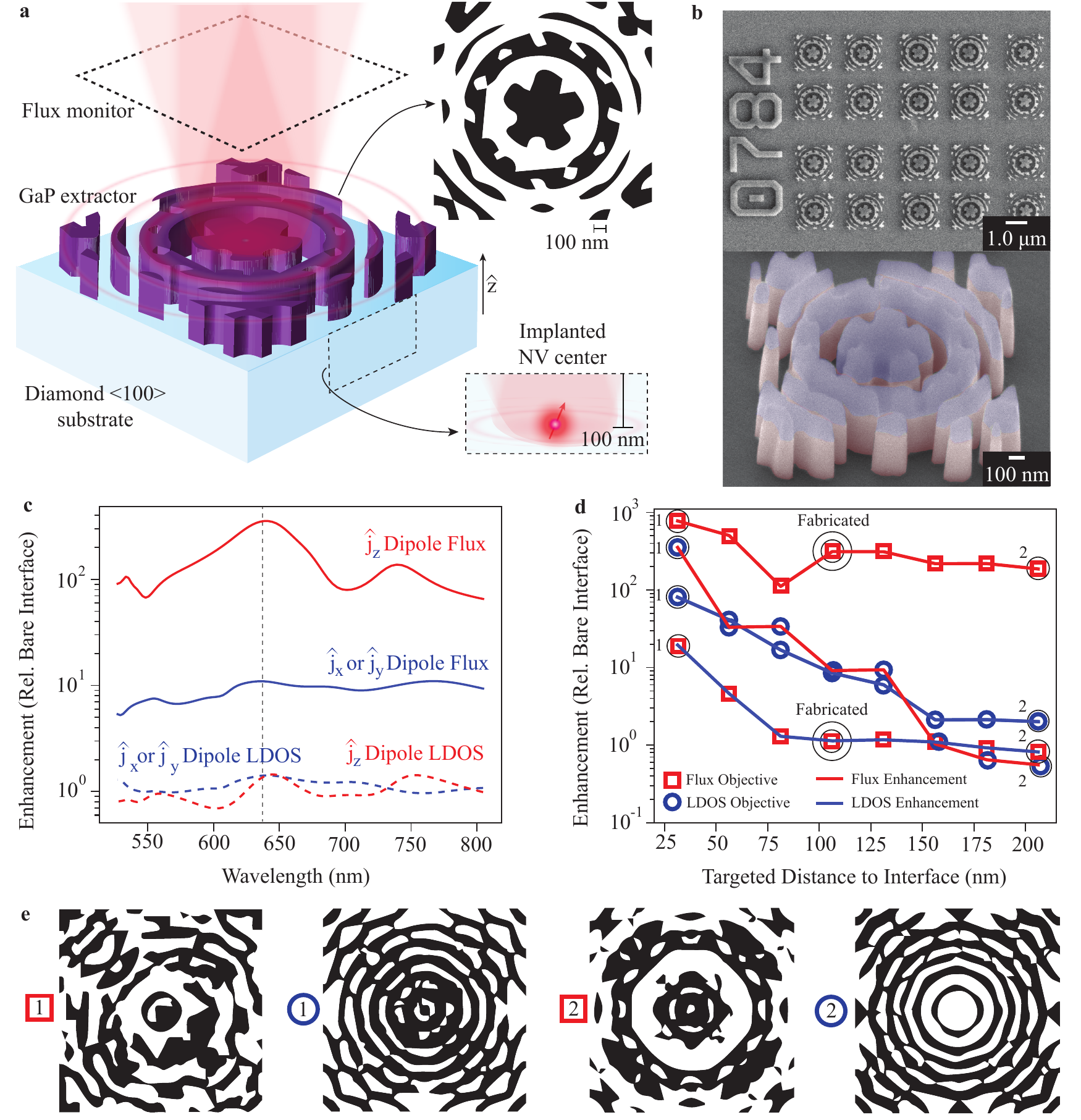}
\caption{\textbf{a.} Schematic of the photon extractor designed via topology optimization (TO).\red{ \textbf{b.} Top: SEM image of an array of fabricated inverse design extractor structures. Bottom: colorized SEM of a single device indicating the different layers; HSQ resist (blue), GaP (pink) and diamond (grey). \textbf{c.} Spectra of LDOS and collection flux enhancement relative to a bare diamond-air interface. Vertical dotted line indicates NV$^-$ ZPL (637\,nm).  \textbf{d.} Performance of TO designed devices targeting z-polarized dipoles at different depths with either LDOS or collection flux enhancement as the optimization objective. Blue circles are LDOS optimized devices and red squares are flux optimized devices; on the different curves, the same symbols at the same target distance correspond to the same device. \textbf{e.} Inverse design structures generated by TO for specific combinations of optimization objective and dipole depth indicated in (d). Around the ZPL, the LDOS optimized structure for dipole depth of 30 nm (1, blue circle) has a dominant mode with $Q=240$; all the other structures have an estimated $Q\approx50$. } } 
\label{fig:fabrication}
\end{figure*}

\section{Design of inverse-designed photon extractors}

The photon extractors are designed for robust enhancement of ZPL photon collection from near-surface NV centers, and modeled via a 3D finite-difference frequency-domain (FDFD) solver, with the frequency set at the negatively-charged NV$^-$ ZPL of 637\,nm. Topology optimization \cite{molesky_inverse_2018} was used to design the device within a design region situated directly on top of the diamond substrate with dimensions 1.5\,\textmu m~$\times$~1.5\,\textmu m~$\times$~0.25\,\textmu m. A harmonic dipole source representing the NV center is situated 100\,nm below the diamond-GaP interface.

\begin{figure*}[ht]
\centering
\includegraphics[width=\textwidth]{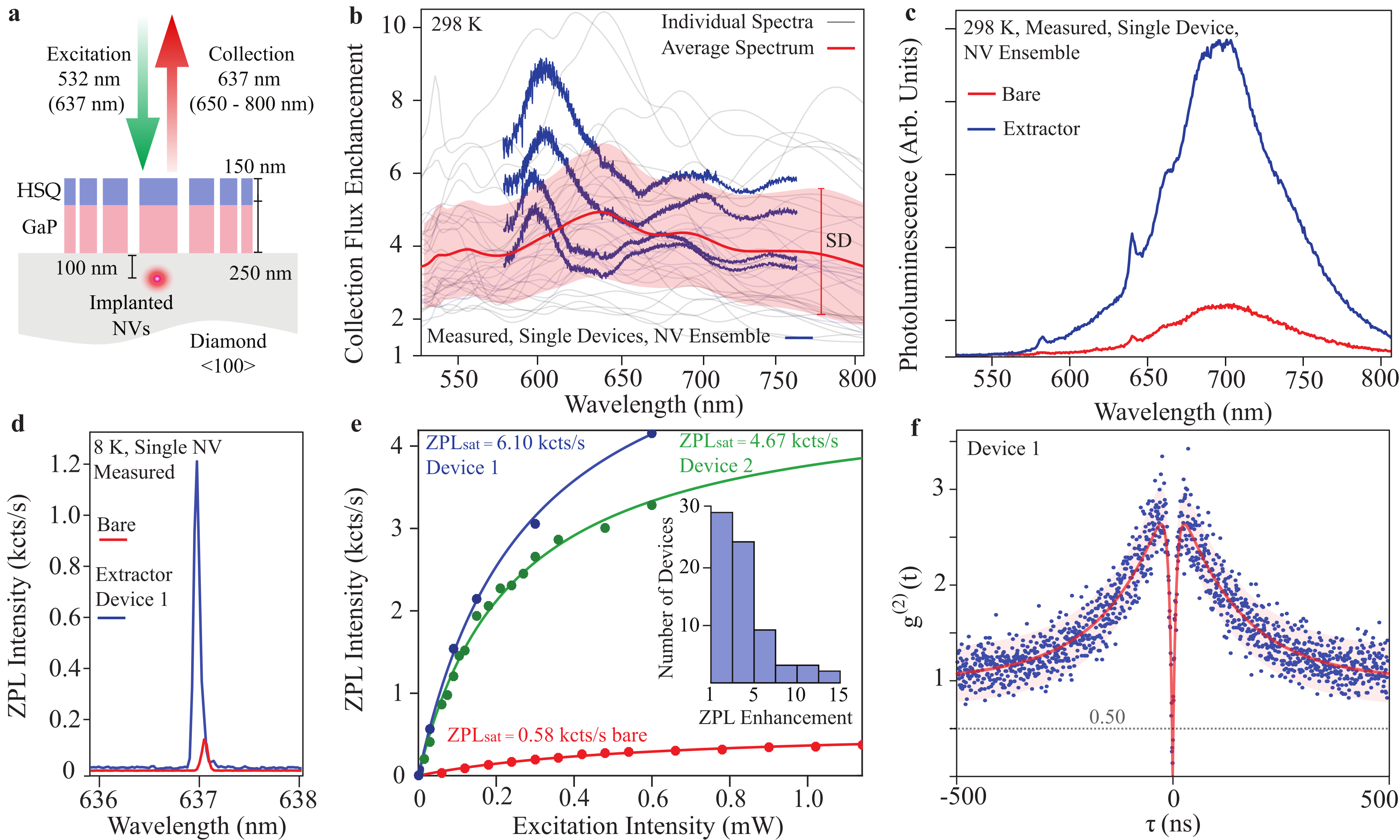}
\caption{\sri{\textbf{a.} Illustration of measurement configuration with implanted NV centers coupled to extractor devices.} \textbf{b.} Comparison of simulated (grey) and measured (blue) broadband enhancement for devices coupled to NV ensemble (Sample A). The red line indicates the average simulated spectrum and the red envelope is one standard deviation around the average. Variation in individual spectra for both simulation and measurement stems from randomness of NV positions relative to any given device.  \textbf{c.}\sri{The broadband PL emission by NV ensemble from an extractor device (blue) vs. the bare-diamond interface (red)}. \textbf{d.}\sri{Single NV ZPL emission on sample B from an extractor device (blue) vs. the bare-diamond interface (red) \textbf{e.} Saturation count rates from devices coupled to single NV centers}. Inset: Histogram of measured devices and the observed ZPL enhancement (Sample B). \textbf{f.} $g^{2}(\tau)$ measurement of device coupled to single NV (Sample B) under 532\,nm 300\,\textmu W excitation.}
\label{fig:spectra_data}
\end{figure*}

The optimization objective for any given dipole polarization is the net Poynting flux through a square collection surface $\partial F$ of sidelength 1.5\,\textmu m, situated 0.4\,\textmu m above the top surface of the device. To make the device robust with respect to uncertainty in the NV center polarization state, a minimax formulation was used so as to maximize the minimum Poynting flux contribution from dipoles with $\unitv{x}$, $\unitv{y}$, and $\unitv{z}$ polarizations. The optimization problem is formulated as follows:
\begin{align}
&\underset{\epsilon_i}{\text{maximize}}& & t \nonumber \\
&\text{subject to}& &t < \min_{j} \left[ \iint_F \vb{S}^j \cdot \unitv{z} \,\, \dd x \dd y \right], \\
&&& \epsilon_i \in [1, \epsilon_{\text{GaP}}] \nonumber
\end{align}

\noindent where $\vb{S}=\frac{1}{2}\Re\{\vb{E}^* \cross \vb{H}\}$ is the Poynting vector, $\epsilon_i$ the permittivity at the $i$th location within the design region, and the superscript $j \in [x,y,z]$ runs over different polarizations of the time harmonic dipole source. $t$ is an auxiliary variable representing the minimum of the flux for the different source polarizations, introduced so the optimization objective and constraints are all differentiable with respect to the degree of freedom. In principle, topology optimization allows the relative permittivity of each spatial discretization voxel within the design region to be a degree of freedom; to accommodate fabrication using electron beam lithography, the actual degrees of freedom form a 2D grid and represent a top-down view of the device. Density and binarization filters \cite{sigmund_morphology-based_2007} were used to restrict the minimum feature size to around 50\,nm, well within the capabilities of electron beam lithography. The adjoint method \cite{giles_adjoint_2000, lalau-keraly_adjoint_2013} is used to efficiently calculate the gradient for the design objective with respect to every degree of freedom.

A schematic representation and $xy$ cross-section of the fabricated design are shown in Fig.~\ref{fig:fabrication}a with the fabricated devices shown in Fig.~\ref{fig:fabrication}b. Fig.~\ref{fig:fabrication}c shows the LDOS and collection flux enhancement spectra of the device for different dipole polarizations, with $\hat{x}$ and $\hat{y}$ polarizations producing nearly the same spectrum due to the resulting near mirror device symmetry. A few salient features of the optimized design are worth commenting upon. To begin with, the extractor yields orders of magnitude larger flux enhancements for $\hat{z}$ compared to in-plane polarized dipoles. Such a vast difference in relative improvement follows from the fact that the bulk of the radiation from a $\hat{z}$ polarized dipole underneath a bare diamond-air interface undergoes total internal reflection, suppressing the collection efficiency by more than a order of magnitude compared to that of an in-plane polarized dipole; the main role of the extractor is therefore to out-couple such previously trapped radiation. Further, device performance is robust against spatial displacement of the \red{emitter}; the enhancement factor decreases by less than 50\% for spatial displacements smaller than 70\,nm from the original location of the NV center \red{(see Fig. SI.1, Supplement 1)}.

\red{Simulating an experimental setup with microscope NA of 0.7, the emitter being an NV center in (100) terminated diamond at a depth of 100 nm coupled to the extractor patterned in GaP (including the resist mask), the fabricated design (Fig.~\ref{fig:fabrication}a) achieves a total collection flux enhancement of 15.7 relative to a bare interface. This is attributed to a 1.4 times local density of states (LDOS) enhancement and a 11 times collection efficiency enhancement (Fig.~\ref{fig:fabrication}c). For comparison, a full GaP slab with the same thickness as the device achieves a 1.2 times LDOS enhancement but actually reduces collection efficiency by 15\% leading to roughly the same collected flux as that of the bare interface. With regards to the absolute collection efficiency, 1.8\% of total NV radiation can be collected from a bare interface, 1.6\% collected with a full GaP slab, and 20\% collected with the optimized design (Section SI.2, Supplement 1 for simulation details and NV flux enhancement spectrum). For calculations of the far field radiation pattern of the extractor device see Fig. SI.3, Supplement 1. }

While the extractor leads to large enhancements in the total collected flux, it does not  significantly increase the net radiation from the NV center, quantified by the LDOS; hence, most of the observed improvement comes from a higher collection efficiency as opposed to fluorescence rate enhancement via the Purcell effect. Fig.~\ref{fig:fabrication}d, e explores this trade-off for a $\hat z$-polarized dipole. \red{At each targeted defect depth, we use TO to design two devices, one with LDOS enhancement as the maximization objective and the other with total collected flux enhancement as the maximization objective.} For shallow dipoles, optimizing for larger flux (square data points) results in a substantial contribution due to LDOS enhancement. However, the divergence between these two design objectives is observed to grow rapidly with increasing dipole depth: beyond a depth of 75\,nm,  devices designed to maximize the collected flux maintain a flux enhancement ratio of about 100, while their LDOS enhancement is close to one. Notably, beyond a dipole depth of 150\,nm, even LDOS optimized devices are only able to achieve LDOS enhancements of roughly two, as the result of the rapidly decaying access of the device to the dipole's near field. Devices that combine the benefits of Purcell enhancement with increased collection efficiency through interference are therefore only needed for NV centers sufficiently close ($\lesssim$ 100\,nm) to the interface. 

\red{Finally, with regards to the device footprint and material choice, we note that for the target emitter depth of 100\,nm for the fabricated design, the device footprint of $(1.5\,$\textmu m)$^2$ covers a large effective NA (relative to the dipole), and TO attempts with larger device footprints show diminishing increase in device performance.} \red{As GaP is a strong dielectric with a high susceptibility, there are limited gains to switching to a higher susceptibility material (Fig. SI.4, Supplement 1), and the low loss of GaP among high susceptibility materials in this frequency range make it particularly attractive for integrated photonics applications \cite{schmidgall_frequency_2018}. }

\section{GaP photon extractors coupled to implanted NVs}

We fabricate photon extractors (Fig.~\ref{fig:fabrication}b) on two samples, one for ensemble-averaged measurements (Sample A) and the other for single-emitter characterization (Sample B). Sample A is a high pressure high temperature synthesised diamond (Element Six, N~$<$~200\,ppm, B~$<$~0.1\,ppm)  implanted with $^{14}$N accelerated to 20~keV and vacuum annealed at 800\,$^\circ$C. Here, annealing forms NVs primarily by vacancy recombination with native nitrogen. The resulting NV distribution is dictated by the vacancy diffusion profile \cite{santori_vertical_2009}, yielding a dense layer ($\approx$100 NVs per 800\,nm excitation spot diameter) of NV centers \sri{$\lesssim$}100\,nm below the diamond surface. Sample B is a chemical vapor deposition diamond (Element Six, electronic grade, N~<~5\,ppb, B~<~1\,ppb) implanted with $^{15}$N accelerated to 85~keV and vacuum annealed at 1200\,$^\circ$C~(see Methods). During annealing, NV centers are formed primarily by vacancy recombination with the implanted nitrogen, yielding a thin layer ($\approx$ 3 NVs per excitation spot) of single NV \sri{centers} 100\,nm$~\pm~$20\,nm from the surface.

After NV-formation, a 250\,nm thick gallium phosphide (GaP) membrane is transferred to each sample via a wet lift-off process~\cite{yablonovitch_van_1990, gould_efficient_2016, schmidgall_frequency_2018}. Electron beam lithography and subsequent plasma RIE etching of the GaP layer forms the 1.5\,\textmu m~$\times$~1.5\,\textmu m photon extractors. The small footprint was chosen for compatibility with on-chip electrodes enabling optical frequency tuning~\cite{acosta_dynamic_2012,schmidgall_frequency_2018}.  Over 100\,000 devices are fabricated in multiple arrays on the 2~mm~$\times$~2~mm diamond substrate. An array of fabricated devices and a false color SEM image highlighting the material stack is shown in Fig.~\ref{fig:fabrication}b. On average, each feature is measured to be within 10\% of the design, with near vertical ($\theta=$~88$^\circ$) GaP sidewalls. \sri{See Section SI.5, Supplement 1 for more details on fabrication.}

\begin{figure*}[t]
\centering
\includegraphics[width=\textwidth]{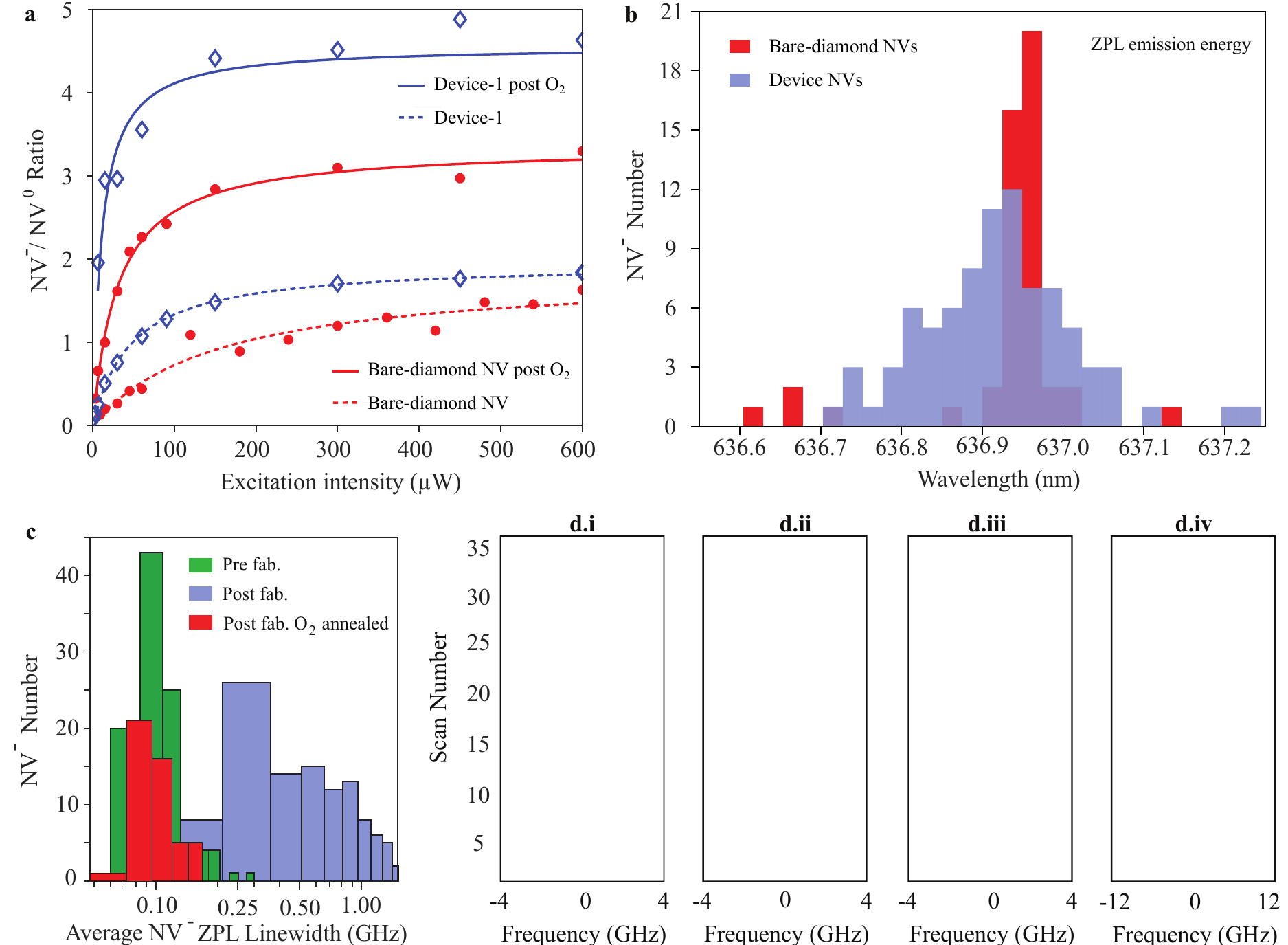}
\caption{(Sample B, T=8\,K) \textbf{a.} The ratio of ZPL intensities of NV$^-$ (637\,nm) and NV$^0$ (575\,nm) charge states for device-coupled and bare-diamond \sri{implanted} NVs as a function of the 532\,nm excitation intensity. \textbf{b.} The distribution of ZPL wavelength for device-coupled and bare-diamond NVs. \textbf{c.} The distribution of  observed average PLE linewidth for bare-diamond NVs. \textbf{d.} PLE characterization of NVs; (d.i) Grown-in deep NV avg. linewidth = 44\,MHz, (d.ii) Implanted NV pre-fabrication avg. linewidth = 66\,MHz, (d.iii) \sri{Bare-diamond} NV post-fabrication after oxygen annealing avg. linewidth = 59\,MHz, (d.iv) Device-coupled NV \sri{after} oxygen annealing avg. linewidth = 844\,MHz. The green markers indicate 532\,nm repump pulses that re-initialize the NV into the NV$^-$ charge state.}
\label{fig:ple_data}
\end{figure*}

\section{Results and discussion}
\subsection{Enhanced photon extraction}
To study NV PL enhancement from the fabricated devices, we use spatially-resolved photoluminescence spectroscopy (Section SI.6, Supplement 1) under off-resonant 532\,nm excitation (Fig.~\ref{fig:spectra_data}a). . First, sample A is utilized for broadband characterization of ensemble-averaged enhancement over the entire NV spectrum. \sri{In this measurement, the 532\,nm laser with a spot diameter of $\approx$ 800\,nm is focused on the center of the device, optimized for maximum PL collection.} \red{Due to both the lateral extent of the excitation spot and the depth distribution of the NV ensemble, the NV-device coupling is highly inhomogeneous.} The room temperature spectra from devices are normalized to \sri{the PL from the nearby bare-diamond}, which spatially varies in intensity by $\approx$14\% (Fig. SI.7b, Supplement 1). We observe an average six-fold enhancement that is relatively flat over 575 to 750\,nm and matches \red{reasonably} well to the theoretical 4-fold enhancement (Fig.~\ref{fig:spectra_data}b, c) \red{calculated by averaging over the NV spatial and dipole distributions (for details on the simulation see Section SI.2, Supplemental 1).} 

Having demonstrated broadband NV PL enhancement, we move on to single-NV coupled devices on sample B and characterize the enhancement of the NV zero-phonon line (ZPL) emission. Identification of single-NV coupled devices is performed by comparing the low temperature (8\,K) NV$^-$ ZPL spectra under 532\,nm excitation of devices to nearby \sri{bare-diamond} NV \sri{centers}  (Fig.~\ref{fig:spectra_data}\sri{d}). Enhanced ZPL collection rates (Fig.~\ref{fig:spectra_data}\sri{e}) were exhibited in 74 of 480 tested devices. The distribution of observed enhancement can be seen in Fig.~\ref{fig:spectra_data}e\sri{(inset)}. \red{Enhancement as high as 14-fold can be observed, which is based on the NV saturation intensity to eliminate the effect of excitation polarization. The saturation intensity of bare-diamond NV centers varies by $\approx$8\%(Fig. SI.7d, Supplement 1).} The placement of NVs with respect to the device optical mode is random; coupled device yield can thus be significantly improved by targeted implantation or pattern alignment to registered defect centers~\cite{marseglia_nanofabricated_2011} during fabrication. 

We verify that the observed enhanced PL emission corresponds to single-emitter\sri{s} using an autocorrelation measurement on the NV$^-$ ZPL at 637\,nm. Fig.\,\ref{fig:spectra_data}f shows the normalized coincidences under continuous off-resonant  excitation for a device-coupled NV. The dip in coincidences at 0 time delay is 0.12 $<$ 0.5 and verifies single-photon emission. Autocorrelation measurements on the eight brightest devices confirm single-NV coupling in six of the devices. In addition to the dip at zero time delay, significant bunching is observed in all devices on the 100\,ns timescale. In modeling the autocorrelation curve (red line, Fig.\,\ref{fig:spectra_data}f), we find it is necessary to include both the NV$^-$ singlet state and the NV$^0$ charge state to reproduce the magnitude and timescale of the bunching for a series of power-dependent measurements. Consideration of charge state dynamics is a critical for NV  device performance as discussed further below.  Consistent with theoretical expectations, a significant excited state lifetime reduction (due to an increased LDOS, Fig.\,\ref{fig:fabrication}c, dashed lines) is not necessary to obtain good agreement between the experimental data and model. Details of the modelling procedure and optimized rates are provided in Section SI.8, Supplement 1. 

\subsection{Charge state and spectral stability}
Single defect qubit devices have requirements beyond photon collection efficiency including charge state stability and spectral homogeneity and stability. These properties should be preserved during device integration. 
Our autocorrelation model suggests that rapid charge-state conversion occurs between the neutral (NV$^0$) and negatively-charged (NV$^-$) states. For both sensing and quantum information applications, the NV$^-$ charge state is required, hence we need to minimize ionization into the NV$^0$ charge state. Every NV center, \sri{whether or not it is coupled to a device}, has emission at both NV$^0$ and NV$^-$ ZPL transitions, with the ratio of the ZPL intensities (Fig.~\ref{fig:ple_data}a) determined by the local Fermi-level\cite{fu_conversion_2010, kato_tunable_2013} and excitation intensity\cite{siyushev_optically_2013}. \red{We also observe a broadening in the inhomogeneous NV$^-$ ZPL distribution and an overall blue shift of $30\pm15\,\mathrm{GHz}$ (Fig.~\ref{fig:ple_data}b) for device-coupled NVs vs nearby bare-diamond NVs. These effects can be attributed to local variation in the strain environment\cite{batalov_low_2009,knauer_-situ_2020}. Identical photon emission from different coupled defects are essential for photon-mediated defect qubit entanglement schemes\cite{bernien_two-photon_2012,humphreys_deterministic_2018}. The observed static inhomogeneity can be bridged by Stark tuning\cite{acosta_dynamic_2012,schmidgall_frequency_2018} or quantum frequency conversion techniques~\cite{takesue_erasing_2008, fan_optomechanical_2016}. Characterization of the local strain environment for individual device-coupled NV centers and the prospect of additional inhomogeneous broadening due to strain induced by the GaP-diamond interface at low temperatures (8\,K) is provided in Section SI.9, Supplement 1.}

High resolution photoluminescence excitation (PLE) spectroscopy gives us further insight into the optical coherence (Fig.~\ref{fig:ple_data}c) and temporal spectral stability of individual defects \sri{(Fig.~\ref{fig:ple_data}d)}. In PLE measurements, a narrow-band tunable laser is scanned across the NV$^-$ ZPL while collecting the NV$^-$ phonon-sideband PL (650\,nm to 800\,nm). From the PLE spectra we obtain the average optical linewidth as well as the scan-to-scan variation in the ZPL frequency, indicating spectral diffusion. During PLE we can \sri{sometimes} observe a loss of the NV$^-$ PL signal due to ionization to the NV$^0$ state. When PL is lost, we apply a short 532\,nm repump pulse (0.1~s) between scans to reinitialize into the NV$^-$ charge state \sri{(as indicated by the green markers in Fig.~\ref{fig:ple_data}d)}. Hence, the interval between repump pulses is another indicator of the NV$^-$ charge state stability. 

For a deep, single NV center incorporated during growth, NV$^-$ is the preferred charge state at low excitation power (NV$^-$/NV$^0=$3.4 at 60 \textmu W of 532\,nm excitation). An average linewidth of 44\,MHz is observed (Fig.~\ref{fig:ple_data}d.i). Minimal spectral diffusion is observed between repump pulses, with the NV $^-$ ZPL frequency exhibiting a standard deviation of 48\,MHz, It is a challenge to demonstrate this level of optical stability with device-coupled implanted near-surface NVs. Prior to device fabrication, the implanted single NV centers exhibit an average optical linewidth of $<$\,100\,MHz (Fig.~\ref{fig:ple_data}c, green). Between repump pulses, the standard deviation of the NV $^-$ ZPL frequency is $\sim$100\,MHz, indicating low spectral diffusion. (Fig.~\ref{fig:ple_data}d.ii). The NV$^-$ charge state remains stable, no ionization is observed for multiple scans over ten minutes of measurement time. This data compares favourably with the grown-in NVs.

Post-fabrication, for both device and \sri{bare-diamond} implanted NV centers, the preferred charge state at low excitation power is NV$^0$. In addition to the low NV$^-$/NV$^0$ ratio (Fig.~\ref{fig:ple_data}a dashed lines), we observe broadening of the average single NV$^-$ linewidth (Fig.~\ref{fig:ple_data}c, blue), rapid ionization and large spectral diffusion. We suspect that although our design avoids etching into the diamond, the GaP photonics plasma etch (Ar/Cl$_2$/N$_2$) modifies the surface termination\cite{widmann_f-_2014,tao_single-crystal_2014} of the diamond and introduces new surface charge traps. Encouraged by prior studies\cite{fu_conversion_2010, yamano_charge_2017}, we performed a post-fabrication oxygen anneal of the sample at 400\,$^\circ$C (for details see Section SI.10, Supplement 1). From Fig.~\ref{fig:ple_data}a (solid lines), we see that the surface treatment more than doubles the NV$^-$/NV$^0$ ratio. For the \sri{bare-diamond} NV centers, we recover the pre-fabrication average NV$^-$ linewidth (Fig.~\ref{fig:ple_data}c, red), but the centers retain the larger spectral diffusion (Fig.~\ref{fig:ple_data}d.iii). For device-coupled NV centers (Fig.~\ref{fig:ple_data}d.iV), we measure an average device-coupled NV linewidth of 844\,MHz (1.5\,GHz pre-O$_2$ annealing). This linewidth is $\approx$ 8 times larger than bare-diamond NVs. Further studies are needed to determine if this fast spectral diffusion can be attributed to the GaP-diamond interface or laser-induced due to modification of the excitation intensity profile. One promising avenue for further reduction in fabrication-induced NV instability is to fabricate the GaP device prior and transfer it via a stamping process~\cite{dibos_atomic_2018}. Our inverse-design optimization framework can accommodate additional constraints such as interconnected support structures required by the stamp-transfer fabrication technique.

\section{Conclusion}
\red{This work utilizes inverse-design techniques to optimize the photon collection efficiency from single NV centers under three conditions chosen for future scalability: (1) the NV center is created via implantation, 100\,nm from the surface (2) the diamond defect qubit host is not etched and (3) the device lateral dimensions are 1.5 ~\textmu m. Despite these constraints, the resulting 14-fold enhancement performs similar to related broadband devices in the literature (6 to 20-fold enhancement) such as solid immersion lenses\cite{hadden_strongly_2010,jamali_microscopic_2014}, diamond nanowires\cite{babinec_diamond_2010}, bullseye gratings\cite{li_efficient_2015}, diamond metalenses \cite{huang_monolithic_2019} and parabolic reflectors\cite{wan_efficient_2018}, all of which required  directly etching the diamond via harsh processes that degrade near-surface NV optical properties~\cite{ruf_optically_2019}}. Other designs based on hybrid metal-dielectric gratings / plasmonic resonances may achieve Purcell enhancement along with high directionality  (up to 15-fold enhancement), but they require the NV to be either very close (few nm) to the surface\cite{karamlou_metal-dielectric_2018} or embedded in the device within a nanodiamond\cite{choy_enhanced_2011,livneh_highly_2016,andersen_hybrid_2018}, both extremely challenging environments for the realization of high charge stability and spectral purity required for quantum information applications. \red{We focused on 100\,nm-deep centers due to the ability create optically coherent centers at this depth~\cite{chu_coherent_2014}; however due to the (relatively) large depth, the resulting design predominantly works as a photon extractor with little emission (LDOS) enhancement. A}s near-surface defect qubit engineering advances, our inverse-design platform can readily be used for even higher enhancement of photon emission from shallower defects\red{. Moreover, the platform} provides the design flexibility for integration with emerging technologies such as stamp-transferred devices~\cite{dibos_atomic_2018} \red{that can eliminate NV exposure to nearly all fabrication processes to further preserve defect qubit coherence}. 

During the preparation of this manuscript we were made aware of a contemporaneous theoretical proposal by Wambold et al.\cite{wambold_adjoint-optimized_2020}. We see the two schemes as complementary, both showcasing the ability of nanophotonic inverse design for handling non-trivial design objectives and constraints. Our design is geared towards quantum information applications, while Wambold {\it et al.} is focused on quantum metrology. 


\section*{Acknowledgments}
This material is based upon work supported by the National Science Foundation under Grants EFMA-1640986 (photonic design and fabrication) and ECCS-1807566 (NV synthesis and characterization). The photonic devices were fabricated at the Washington Nanofabrication Facility, a National Nanotechnology Coordinated Infrastructure (NNCI) site at the University of Washington which is supported in part by funds from the National Science Foundation (awards NNCI-1542101, 1337840 and 0335765). We thank E.R. Schmidgall for help with fabrication process development and pre-implantation etching of our diamond samples. 

\section*{Disclosures}
The authors declare no conflicts of interest.
\\

\noindent See Supplement 1 for supporting content.

\section{References}

\bibliography{sri_references,AdditionalRefs}

\end{document}


\title{Supplementary Information: Inverse design photon extractors for optically addressable defect qubits}
\author{Srivatsa Chakravarthi}
\email{srivatsa@uw.edu}
\affiliation{Department of Electrical and Computer Engineering, University of Washington, Seattle, Washington 98195, USA}
\author{Pengning Chao}
\email{pengning@princeton.edu}
\affiliation{Department of Electrical Engineering, Princeton University, Princeton, NJ, USA}
\author{Christian Pederson}
\affiliation{Department of Physics, University of Washington, Seattle, Washington 98195, USA}
\author{Sean Molesky}
\affiliation{Department of Electrical Engineering, Princeton University, Princeton, NJ, USA}
\author{Andrew Ivanov}
\affiliation{Department of Physics, University of Washington, Seattle, Washington 98195, USA}
\author{Karine Hestroffer}
\affiliation{Department of Physics, Humboldt-Universitat zu Berlin, Newtonstrasse, Berlin, Germany}
\author{Fariba Hatami}
\affiliation{Department of Physics, Humboldt-Universitat zu Berlin, Newtonstrasse, Berlin, Germany}
\author{Alejandro W. Rodriguez}
\affiliation{Department of Electrical Engineering, Princeton University, Princeton, NJ, USA}
\author{Kai-Mei C Fu}
\affiliation{Department of Electrical and Computer Engineering, University of Washington, Seattle, Washington 98195, USA}
\affiliation{Department of Physics, University of Washington, Seattle, Washington 98195, USA}
\date{\today}
\maketitle            

 
 \newpage
 
\section{Robustness to Shifts in Dipole Position}

\begin{figure*}[ht]
\centering
\includegraphics[width=6in]{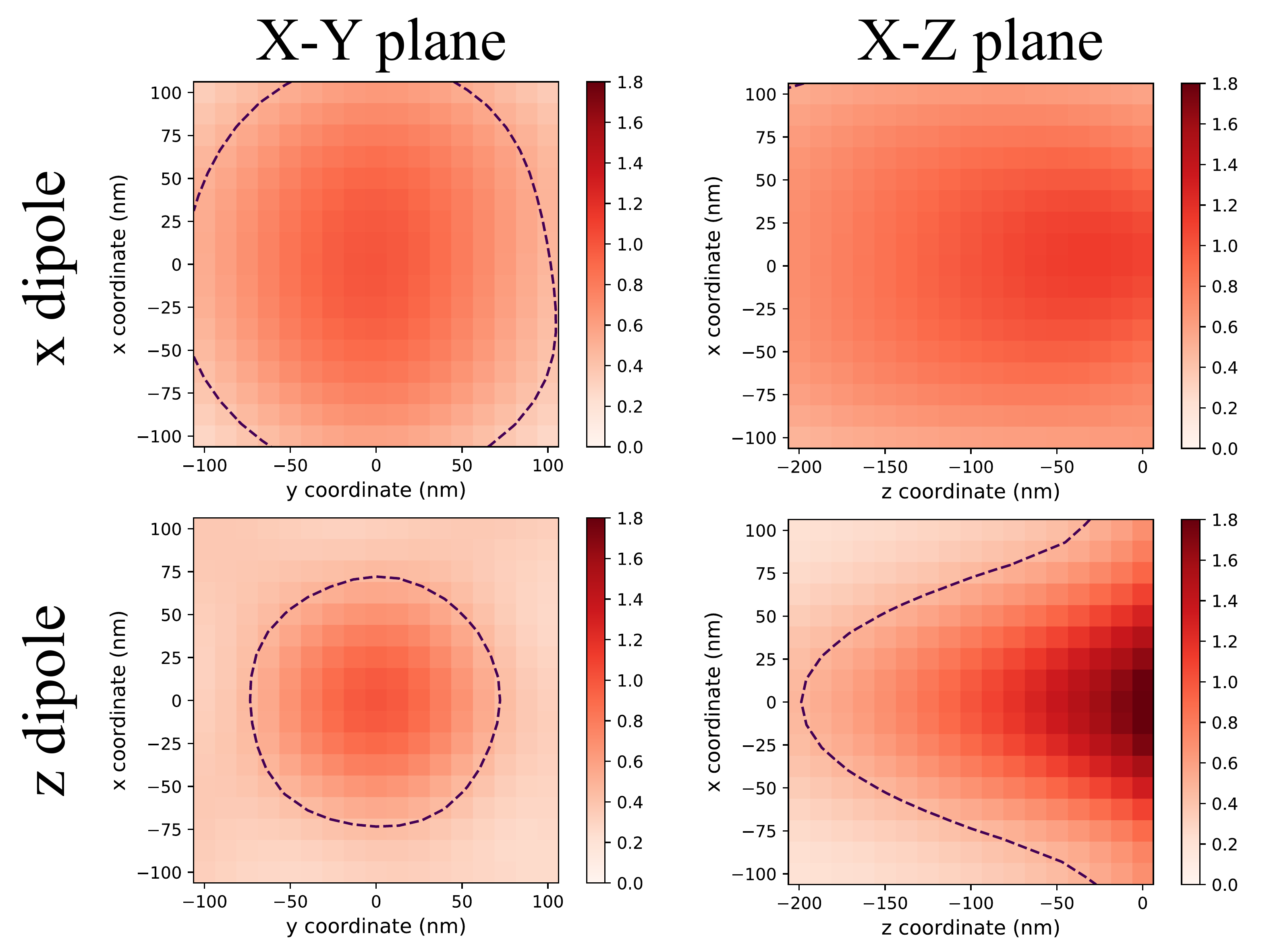}
\caption{Change in collection flux enhancement for x/z polarized dipoles shifting from inverse design position. Coordinate origin is at the center of the bottom face of the device, inverse design position situated at $(0,0,-100)$\,nm. Enhancement factors are normalized to that at the inverse design position; dashed lines indicate contour where enhancement is 50\% of that at inverse design position. There is an increase in enhancement factor as the dipole shifts closer to the surface. }
\label{fig:SI_8}
\end{figure*}

\newpage
\section{Numerical Calculation Details and Average NV Flux Enhancement Spectrum}
Topology optimization was done using in-house finite-different frequency-domain code. Spectra calculations were performed using the open source FDTD package MEEP \cite{oskooi_meep_2010}. 

\begin{figure*}[!htb]
\centering
\includegraphics[width=4.5in]{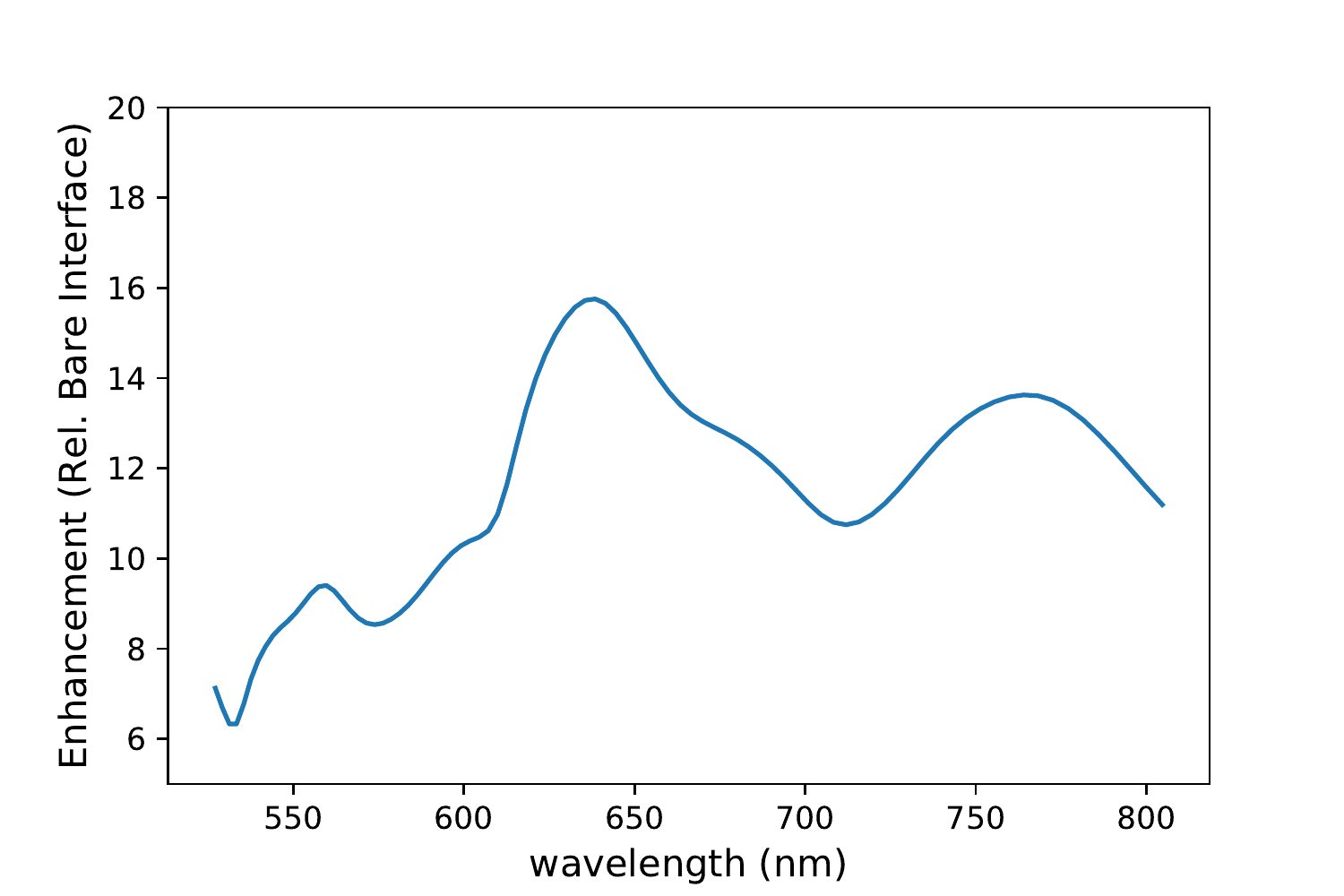}
\caption{Average flux enhancement spectrum for an NV emitter in (100) terminated diamond. }
\label{fig:SI_NV_flux_spectrum}
\end{figure*}

To obtain spectrum results for an actual single NV center embedded in (100) terminated diamond with two cross-polarized dipole moments lying in the (111) plane, we take into account both the effect of local strain on the dipole polarization and the preferential excitement of one dipole moment during measurement. Local strain is modeled as giving a random orientation to the two cross-polarized dipole moments within the (111) plane, and in numerical calculations we sample uniformly the possible orientations to obtain an average. For a given orientation, the dipole moment that has the larger upwards collected flux under a bare diamond-air interface is preferentially excited; we set the dipole moment used in the numerical calculation accordingly. The results are shown in Figure \ref{fig:SI_NV_flux_spectrum}.

Spectrum results in Figure 2b for an NV ensemble are simulated by sampling the positions of a collection of NV centers according to the density distribution of the ensemble. The polarization and phase for each NV center is picked uniformly at random for simplicity. The illuminating laser is modeled as a Gaussian beam, and the incident field within the diamond is calculated beforehand using MEEP. The amplitude of the dipole for each NV center is set proportional to the local incident field along the dipole direction. The flux spectrum for the given configuration is then calculated. The average of the calculations of many randomly sampled configurations produces the red curve in 2b. 

\section{Far Field Radiation Pattern}
\begin{figure*}[h]
\centering
\includegraphics[width=5.5in]{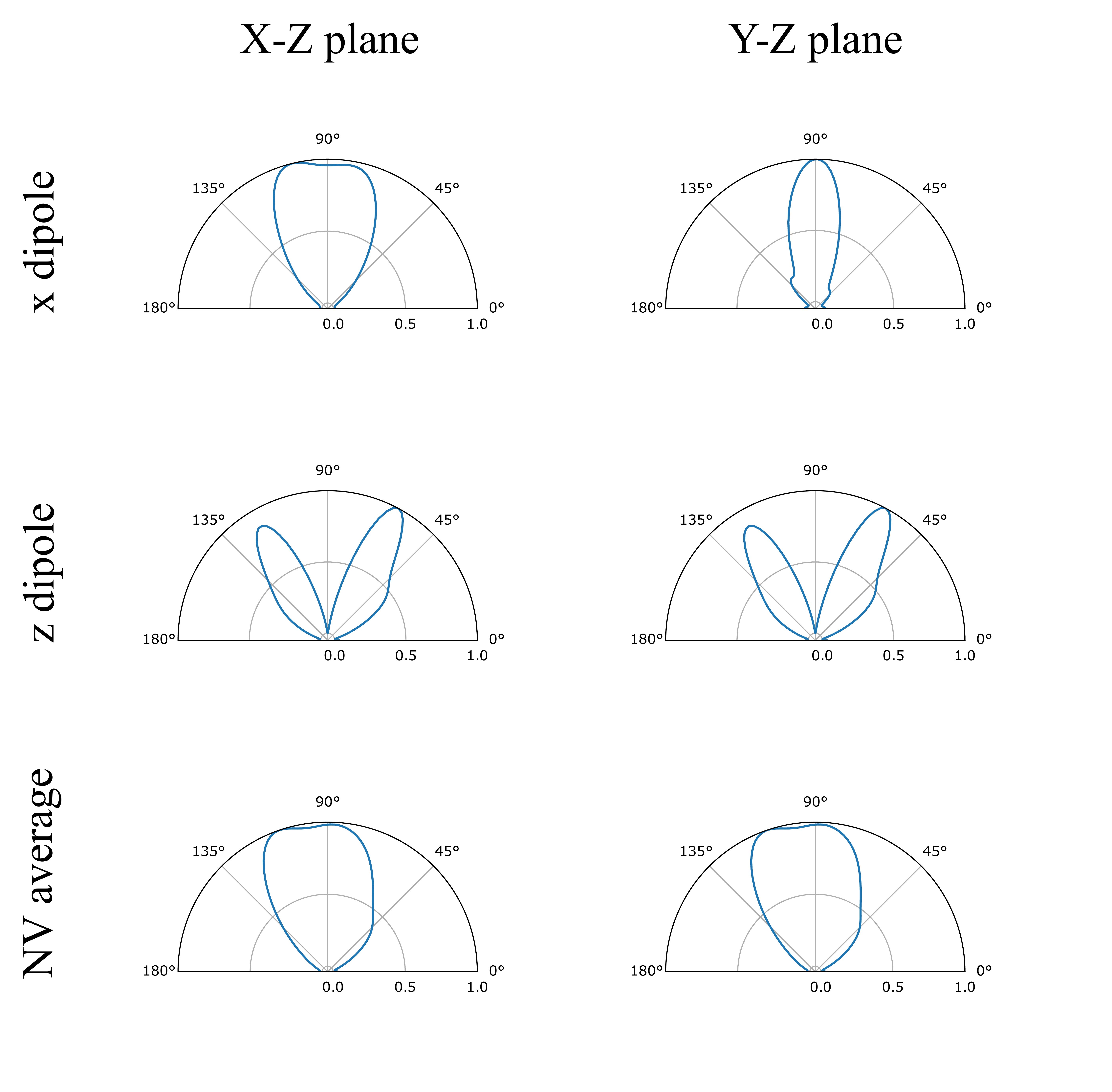}
\caption{Far field radiation pattern in the center x-z and y-z planes for a x-polarized dipole, z-polarized dipole, and average NV dipole, respectively, all at a depth 100 nm from the device. The radiation patterns for the z-polarized dipole and the average NV dipole are symmetric with respect to the y=x plane, due to the underlying symmetry of the design. For the same reason, the radiation pattern of the y-polarized dipole is the mirror image of that of the x-polarized dipole.}
\label{fig:far_field_pattern}
\end{figure*}

Far field radiation patterns for emitters situated 100 nm below the interface at the center of the device are calculated using MEEP and shown in Figure \ref{fig:far_field_pattern}. The average pattern for an NV emitter in (100) terminated diamond is calculated via the procedure outlined in section SI2.
 
 \section{Topology Optimization Results for Varying Material Susceptibility}
\begin{figure*}[ht]
\centering
\includegraphics[width=4.5in]{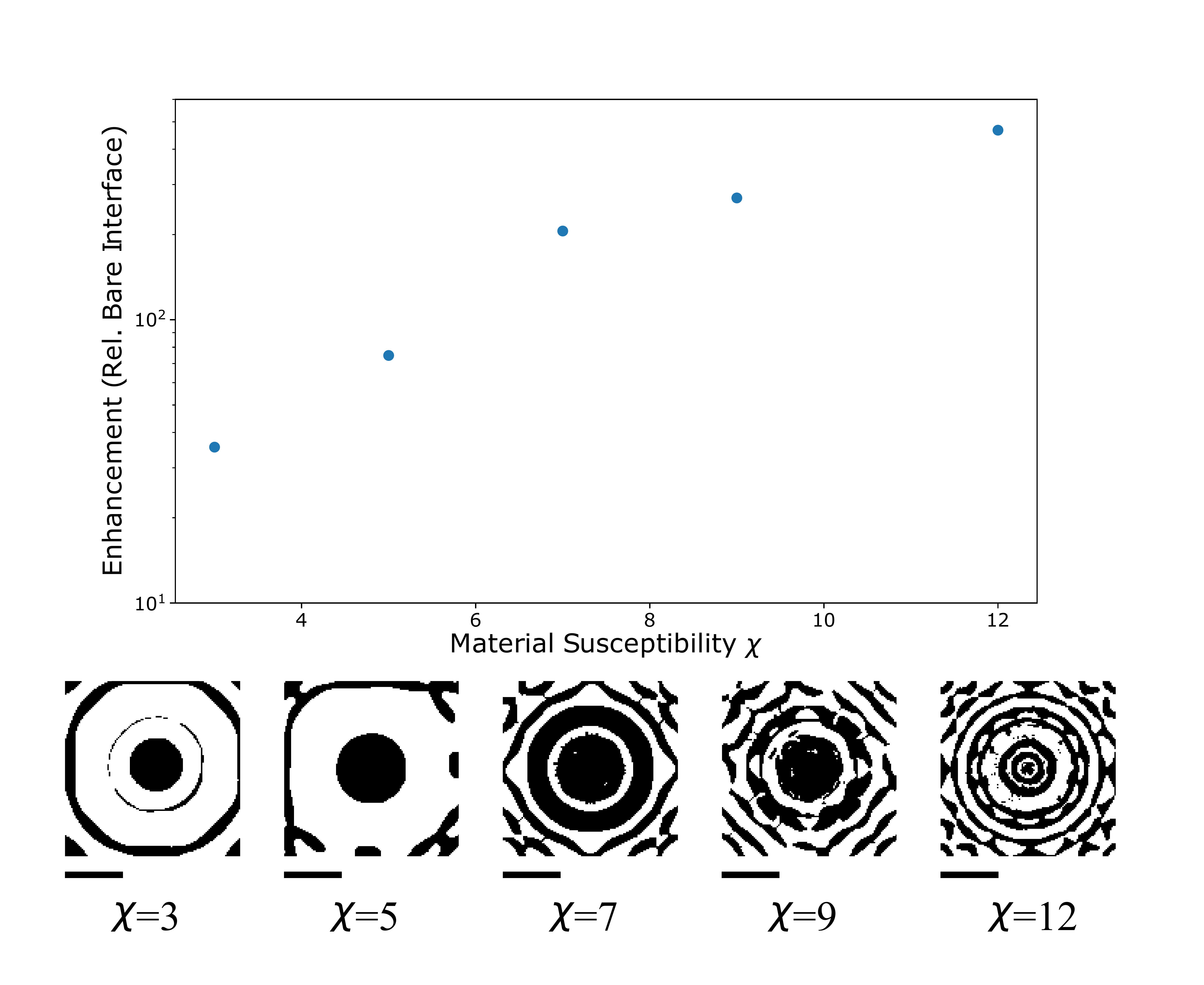}
\caption{Topology optimization results of collection flux enhancement for a z-polarized dipole at different material susceptibility, with cross-sections of the designs. Scale bars are 500 nm.}
\label{fig:SI_10}
\end{figure*}

\newpage
\section{Fabricated devices}  

\begin{figure*}[h]
\centering
\includegraphics[width=6in]{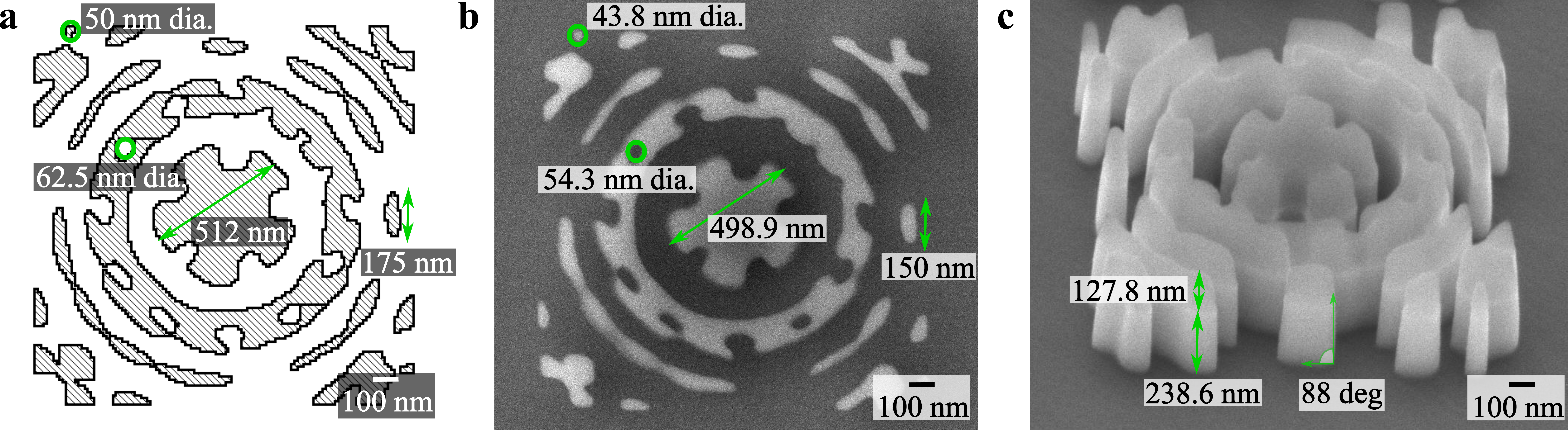}
\caption{\textbf{a.} Design of the photon extractor. \textbf{b.} Top down SEM image of the photon extractor pattern (before GaP plasma etch) with annotated measurements (Sample B). \textbf{c.} SEM image of the photon extractors (after the GaP plasma etch) with annotated layer thicknesses and sidewall angle (Sample B).}
\label{fig:SI_2}
\end{figure*}

Electron beam lithography (100\,kV, 2\,nA, JEOL-6300) is utilized to pattern the designed photon extractor to a thin (150\,nm) HSQ resist layer on top of the GaP-on-diamond stack. However, charging can have a severe impact on the patterning resolution  because of the non-conductive diamond substrate even with the thin GaP membrane. To mitigate charging, we spin a water soluble polymer (4.5\% PSS + 1\% TritonX) over the HSQ resist layer and sputter $\sim$ 5\,nm of Au+Pd metal. This metal layer readily lifts-off during development of the HSQ resist. Multiple arrays of devices are patterned with the design feature size and (x,y) aspect ratio are varying in 12.5\,nm increments across the array to compensate for fabrication variations.The fabricated devices were observed under scanning electron microscopy for process validation and determination of device yield. It can be seen in Fig.~\ref{fig:SI_2} that on average each feature is within 10\% of the designed dimensions.    

\vspace{3em}
\section{Optical measurements}
    
\begin{figure*}[h]
\centering
\includegraphics[width=6in]{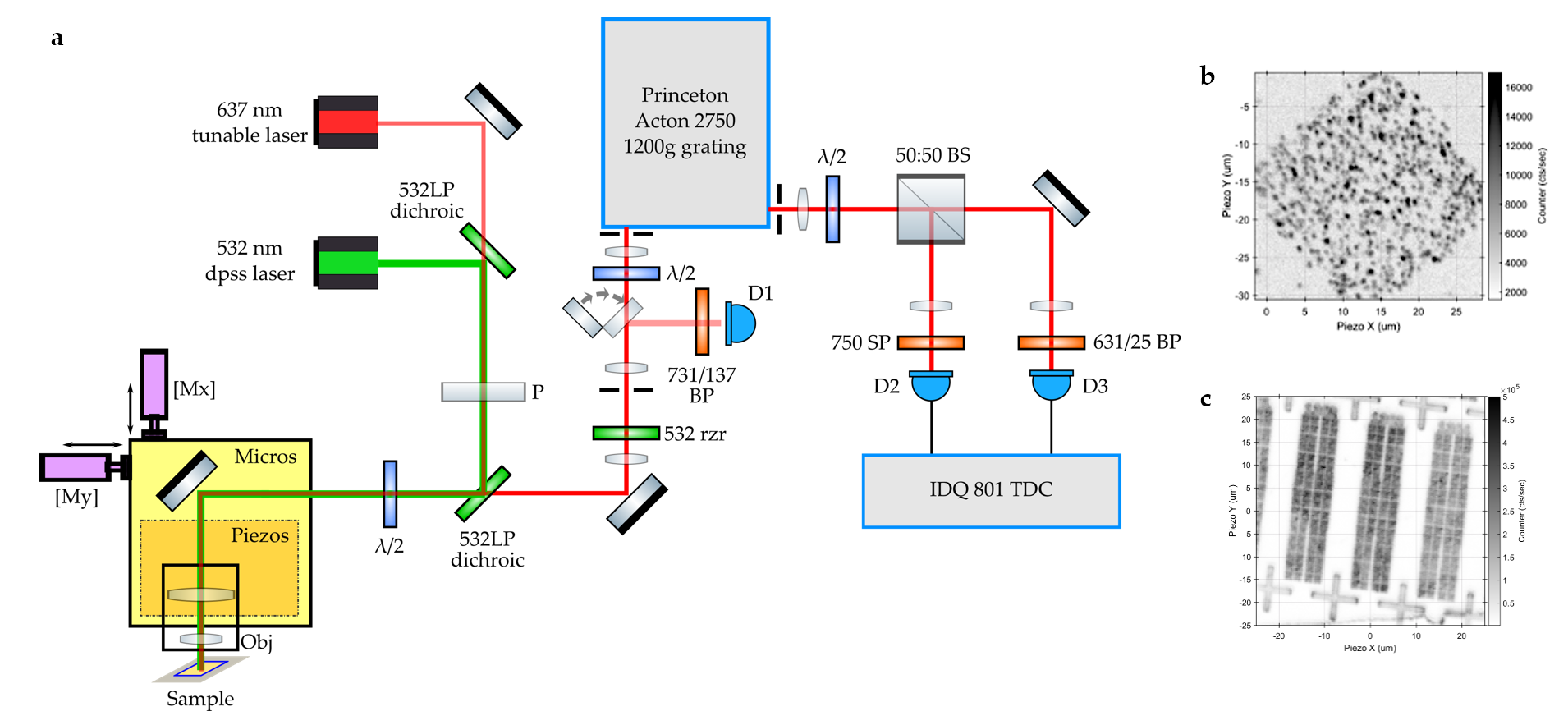}
\caption{\textbf{a.} Illustration of full microscope setup. \textbf{b.} Confocal image of NV PL from an implanted square outside the grid of photonics devices on sample B. \textbf{c.} Confocal image of PL from a grid of photonics devices on sample B.}
\label{fig:SI_1}
\end{figure*}

Optical measurements are performed using a home-built confocal microscope (Fig.~\ref{fig:SI_1}a) with a 0.7 NA, 60x objective (Nikon plan-fluor 60X). A piezo stage (Piezo systems-Jena) is used for fine positioning and confocal imaging of 80x80\,\textmu m$^2$ areas (see Fig.~\ref{fig:SI_1}b, c), while a micrometer stage comprised of two automated actuators (Newport TRA25CC) is used for coarse positioning. The NV centers are optically excited with a 532\,nm diode-pumped solid-state laser (RGBLase LLC FB-532) at powers between 0.1-5\,mW with a spot diameter of $\sim$800\,nm. A polarizing beamsplitter with an automated half-wave plate is used in the excitation path to preferentially excite a given NV orientation. For off-resonant measurements, a combination of dichroic (Semrock 532LP) and 532\,nm razor edge filter is used to attenuate the reflected excitation signal. Spectra are taken with a Princeton Acton 2750 spectrometer with either a 300, 1200 or 1800 groove grating. For resonant excitation studies a New Focus Velocity tunable 637\,nm external-cavity diode laser reflected from a 90/10 T/R beam sampler is used in the excitation path. The phonon-sideband (PSB) emission is filtered in the collection path by a combination of 532\,nm razor edge and a 660-800\,nm bandpass filter (Semrock FF01-731/137-25). The NV$^-$ PSB photons are measured with a fiber coupled avalanche photodiode (Excelitas SPCM-AQ4C). Photoluminesence from Sample A is attenuated with an optical density filter (Newport-530-OD1) to avoid detector saturation.\\

On sample B, post-fabrication, background fluorescence is observed under 532\,nm excitation from the resist (HSQ) and GaP structure. To mitigate this issue only the ZPL is utilized for off-resonant optical characterization. For the photon autocorrelation measurements, the NV$^-$ ZPL photons are filtered through the spectrometer with the 1200 groove grating. The filtered photons are then incident on a 50/50 plate beam splitter, with the two output modes measured by free-space coupled low dark count ($<$30 cts/s) avalanche photodiodes (Excelitas SPCM-AQRH-16). Two additional filters (Semrock 750SP, 631/25BP) are introduced to reduce spurious counts from detector afterpulsing. The photons counts are recorded on two independent channels by a timing board (ID Quantique 801 TDC).\\

During resonant excitation measurements, if the NV center under study is in a high strain environment (evident by the splitting of the $E_x$ and $E_y$ excited state transitions under off-resonant excitation), sidebands at 2.88\,GHz are added to the resonant scanning laser using a fiber-coupled electro-optic modulator (EOSPACE PM-0K1-PFA-637). 
\hfill \\

\newpage

\section{NV PL intensity and spatial uniformity}    
\begin{figure*}[h]
\vspace{1em}
\centering
\includegraphics[width=4.8in]{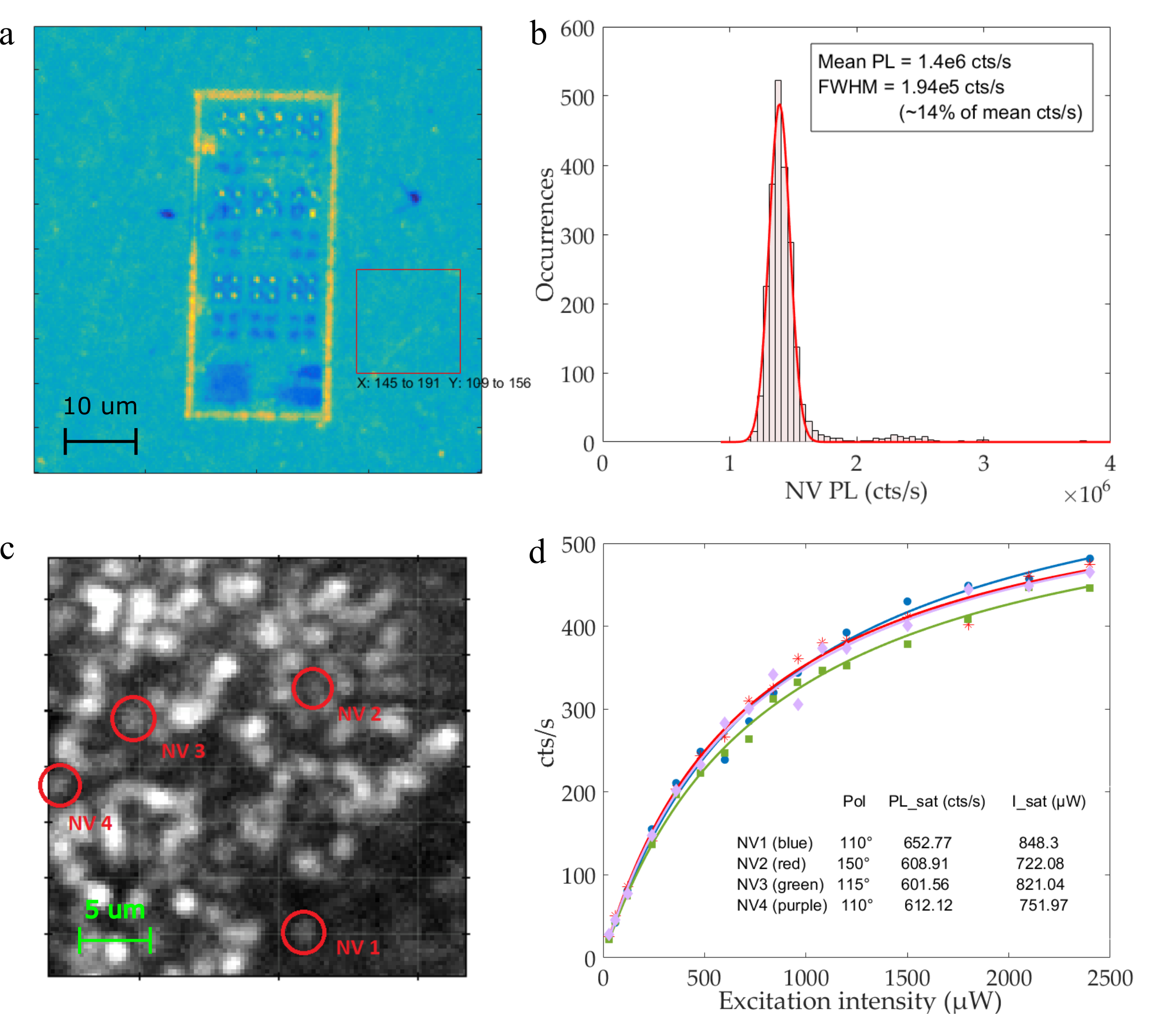}
\caption{\textbf{a.} Confocal PL image of Sample A. \textbf{b.} Histogram of the the PL intensity in the red square shown in (a), when normalized to the PL from a single NV center, we obtain a mean NV density estimate of $\approx$ 100 NV centers per 800\,nm excitation spot. \textbf{c.} Confocal image of an implantation region on sample B showing individual NV centers. \textbf{d.} Measured NV PL intensity vs. optical excitation intensity for the NV centers shown in (c). The excitation polarization for each NV is optimized for maximum PL intensity. }
\label{fig:SI_1b}
\end{figure*}

The implanted ensemble NV density for sample A is estimated by measuring the NV PL distribution over a 10\textmu m $\times$ 10\textmu m confocal scan area (shown as the red box in Fig.~\ref{fig:SI_1b}(a)) on the surface of the diamond sample. The observed NV PL distribution for this region is shown in Fig.~\ref{fig:SI_1b}(b), here the variation in intensity (i.e. the spatial uniformity of the NV density) is estimated as the FWHM of a normal distribution. Normalizing the Ensemble PL by the the PL from a single NV (under similar excitation conditions), gives us a mean NV density of 101 NVs/excitation spot (with a SD = 7 NVs/spot).\\

On sample B, to quantify the PL intensity of single NVs, we perform a confocal scan of an implantation region (Fig.~\ref{fig:SI_1b}(c)) and measure the PL saturation behavior for a few spatially isolated NV centers. The observed NV PL counts at saturation are highly uniform with a spread of just 8.2\% (Fig.~\ref{fig:SI_1b}(d)).    

\newpage
\section{Model for Autocorrelation ($g^{(2)}(t)$) Measurements}

\begin{figure*}[h]
\centering
\includegraphics[width=5.5in]{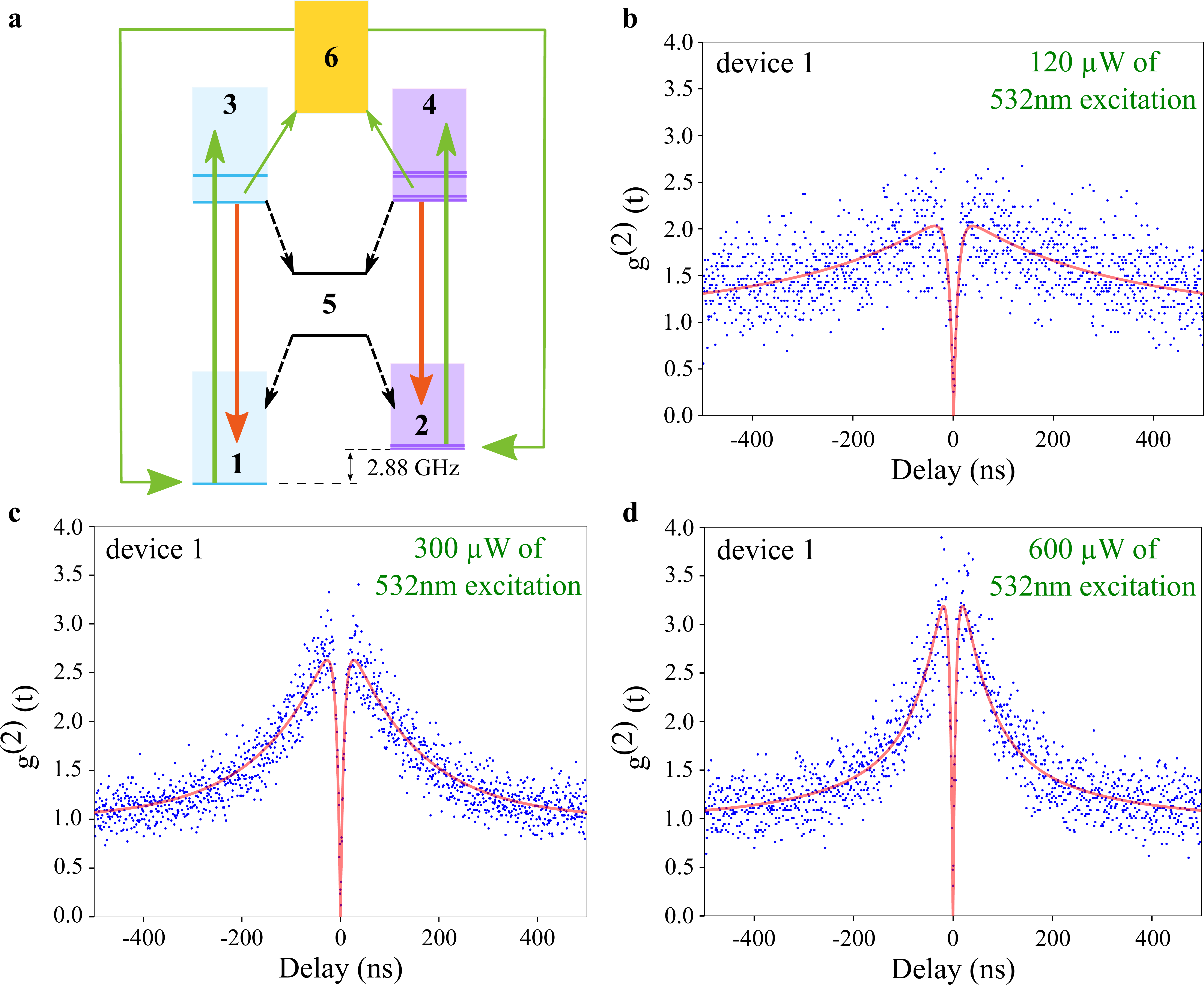}
\caption{\textbf{a.} Energy levels and transitions of the NV$^-$ and NV$^0$ charge states utilized for modelling the g$^{(2)}$($\tau$) data. \textbf{b.} g$^{(2)}$($\tau$) for 120\,\textmu W non-resonant excitation. \textbf{c.} g$^{(2)}$($\tau$) for 300\,\textmu W off-resonant excitation. \textbf{d.} g$^{(2)}$($\tau$) for 600\,\textmu W non-resonant excitation. Note the increase in bunching peaks with increase in excitation intensity.}
\label{fig:g2}
\end{figure*}

The normalized autocorrelation measurements on the NV$^-$ ZPL photons emitted shown in Fig.~\ref{fig:g2}(b-d) provide definitive evidence that we are observing enhancement from a single quantum emitter, yet also reveal significant photon bunching. Power dependent photon bunching is a well-documented effect in NV centers, however the number of coincidences we observed at delays around 20\,ns for the 300\,$\mu W$ and 600\,$\mu W$ data is larger than typically observed in literature. We observe similar bunching in implanted, bare-diamond coupled NV centers in the same sample, but under higher excitation intensity. This similarity suggests that the devices enhance the excitation field at the defect, but are not responsible for the presence of the bunching. Photon bunching is detrimental for many applications since it arises from non-cycling transitions shelving the NV center into inaccessible, or dark, states.

The connection between the photon statistics and the NV center's state is given by the following equation, 

\begin{equation}
g^{(2)}(\tau)=\frac{p(e,\tau | g,0)}{p(e,\infty)},
\end{equation}
where $p(e,\tau | g,0)$ is the probability of finding the NV center in the excited state at time delay $\tau$ given that it was in the ground state at 0 time delay, and $p(e,\infty)$ is the steady state population in the excited state under continuous excitation.

We model the NV center as a 6 level system in order to account for the NV$^-$ ground states, excited states, singlet states, and NV$^0$ states as shown in Fig.~\ref{fig:g2}(a). In our labelling scheme, manifolds 1 and 3 corresponds to the $m_s=0$ ground and excited states respectively. Manifolds 2 and 4 correspond to the $m_s=\pm 1$ ground and excited states.  Manifold 5 contains the singlet states, and manifold 6 consists of the NV$^0$ states. The probability of being in level i is $p_i$ and the rate from level i to j is given by $k_{ij}$, and P is the optical excitation power. Coherences between the different energy levels are assumed to decay on time scales faster than the excitation/relaxation rates due to the non-resonant pumping, which simplifies the model to following set of 6 rate equations.
\begin{equation}
\label{eq:master}
\begin{medsize}
\begin{bmatrix}
\dot{p}_{1} \\
\dot{p}_{2} \\
\dot{p}_{3} \\
\dot{p}_{4} \\
\dot{p}_{5} \\
\dot{p}_{6}
\end{bmatrix} =\begin{bmatrix}
-k_{13}(P)&0&k_{31}& 0 & k_{51} & k_{61}(P^2)\\
0 &-k_{13}(P)&0&k_{31}& k_{52}& 2k_{61}(P^2)\\
k_{13}(P)&0&-(k_{31}+k_{36}(P)+k_{35})&0&0&0\\
0&k_{13}(P)&0&-(k_{31}+k_{36}(P)+k_{45})&0&0\\
0&0&k_{35}&k_{45}&-(k_{51}+k_{52})&0\\
0&0&k_{36}(P)&k_{36}(P)&0&-3k_{61}(P^2)\\
\end{bmatrix}
\begin{bmatrix}
p_{1} \\
p_{2} \\
p_{3} \\
p_{4} \\
p_{5} \\
p_{6}
\end{bmatrix}
\end{medsize}
\end{equation}

One subtlety is that each level of our model corresponds with a manifold of NV states. This has implications in our treatment of off-resonant excitation into the excited state, the multiplicity of the $m_s=\pm 1$ level, and the NV$^0$ internal dynamics. We address the issue of off-resonant excitation by taking the excitation rate to depend linearly on power, while the photon emission rate is power independent. We account for the multiplicity of the $m_s=\pm 1$ states by properly normalizing the branching ratios. For example, the recombination from the NV$^0$ manifold into $m_s=\pm 1$ is taken to be twice the recombination into $m_s=0$. Lastly, we ignore the internal dynamics of NV$^0$, and treat recombination as a two-photon process that in the weak excitation limit should scale as the power squared.

The initial condition is determined by the spin polarization, $p_1/(p_1+p_2)$, and is determined by forcing our solution to be self consistent. In other words  $p_1(0)/(p_1(0)+p_2(0))=p_1(\infty)/(p_1(\infty)+p_2(\infty))$. We find that the equilibrium spin polarization approaches a maximum value of 0.8 at low power and decreases with power.

Older models of NV dynamics ignore ionization and recombination, and focus on the role of the NV$^-$ singlet state and spin non-conserving transitions \cite{manson_nitrogen-vacancy_2006}. We are unable to reproduce the power dependent bunching we observe with the rates from these models.
Our model is more similar to recent literature models which account for both the long-lived NV$^-$ singlet state, as well as the NV$^0$ charge state, but we ignore internal dynamics within the singlet and neutral states \cite{meirzada_negative_2018,aude_craik_microwave-assisted_2020} as we are able to obtain satisfactory agreement simulataneously fitting the three power-dependent $g^{(2)}$ curves with the simplified model (Fig.\,\ref{fig:g2}. We begin the optimization with rates taken directly from these models and optimize over each parameter. This is necessary, since the rates can vary significantly between different NV centers (Table~\ref{table:rates}). We find the final optimized rates are in reasonable agreement with reported values, with the only disagreement being in lower shelving rates into the singlet state, and lower deshelving rates out of the singlet states. Nevertheless, it is clear that ionization and recombination are important factors in the emission properties of an NV center, and should be considered in a variety of applications. \\

\begin{table}[]
\centering
\begin{tabular}{|M{2cm}|M{2cm}|M{2cm}|M{2cm}|M{2cm}|M{2cm}|}
\hline
\textbf{\begin{tabular}[c]{@{}c@{}}Rate\\  (\textmu s$^{-1}$)\end{tabular}} & \textbf{\begin{tabular}[c]{@{}c@{}}Model 1\\ NV 1\\\cite{meirzada_negative_2018}\end{tabular}} & \textbf{\begin{tabular}[c]{@{}c@{}}Model 1\\ NV 2\\\cite{meirzada_negative_2018}\end{tabular}} & \textbf{\begin{tabular}[c]{@{}c@{}}Model 2\\ \\\cite{manson_nitrogen-vacancy_2006}\end{tabular}} & \textbf{\begin{tabular}[c]{@{}c@{}}Model 3\\ \\\cite{aude_craik_microwave-assisted_2020}\end{tabular}} & \textbf{This Model} \\[2ex] \hline
\begin{tabular}[c]{@{}c@{}}$k_{13}$\\  (1 mW)\end{tabular} & 10 & 10 & NA & 59 & 82 \\[2ex] \hline
$k_{31}$ & 77 & 66 & 77 & 75 & 89 \\[2ex] \hline
$k_{35}$ & 0  & 7.9 & 0 & 11 & 17 \\[2ex] \hline
$k_{45}$ & 30 & 53 & 30 & 80 & 17 \\[2ex] \hline
$k_{51}$ & 5 & 1.7 & 3.3 & 2.6 & 1.7 \\[2ex] \hline
$k_{52}$ & 0 & 1 & 0 & 2.3 & 0.50 \\[2ex] \hline
\begin{tabular}[c]{@{}c@{}}$k_{36}$\\  (1 mW)\end{tabular} & 852 & 852 & 0 & 2.2 & 24 \\[2ex] \hline
\begin{tabular}[c]{@{}c@{}}$k_{61}$\\  (1 mW)\end{tabular} & 45 & 45 & 0 & 47 & 11 \\[2ex] \hline
\end{tabular}
\caption{Rates for modelling NV dynamics}
\label{table:rates}
\end{table}

\section{Effect of strain at the GaP-diamond interface on the excited state energy structure of implanted NV centers}
\begin{figure*}[h]
\centering
\includegraphics[width=5.5in]{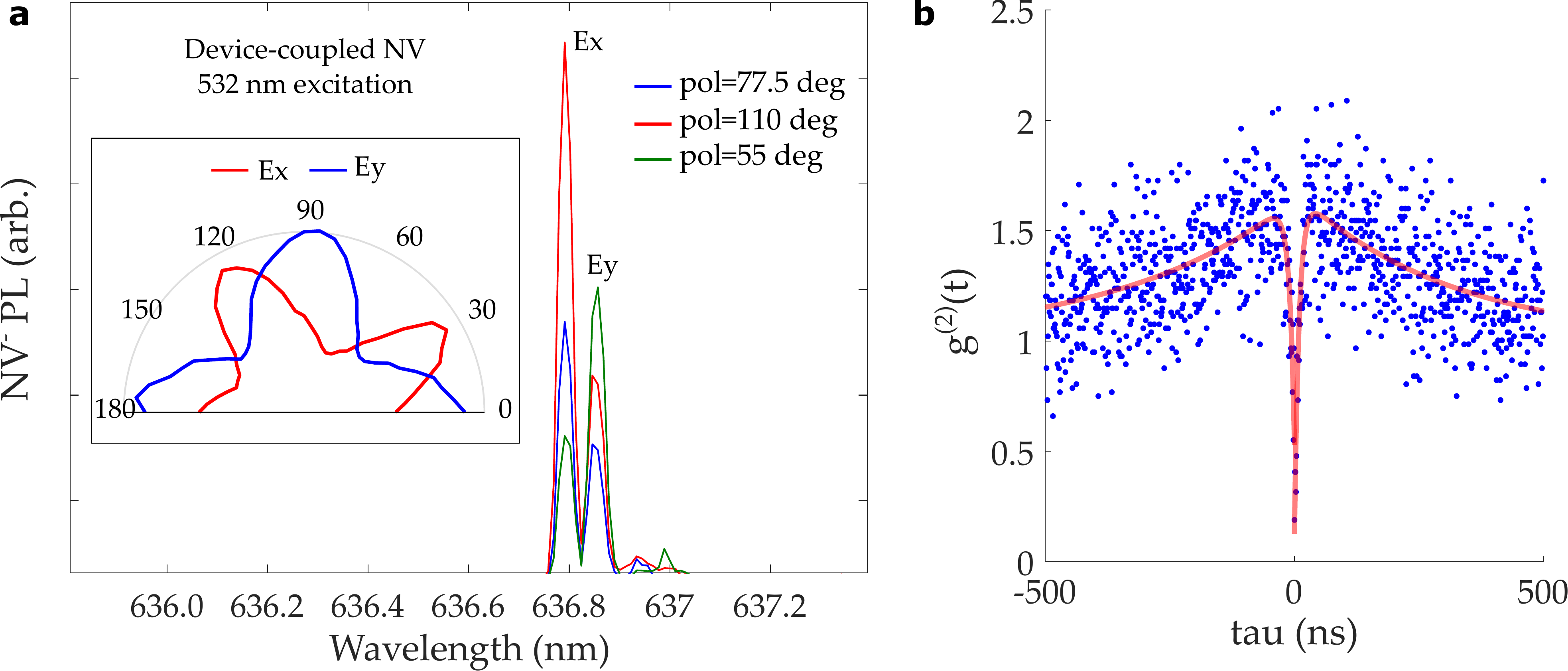}
\caption{\textbf{a.} Spectra of a device-coupled NV center under 532\,nm off-resonant excitation showing a $\sim$48\,GHz (Ex, Ey) splitting. Inset: excitation polarization dependence of the (Ex. Ey) intensity. \textbf{b.} Autocorrelation measurement on the device showing dip of 0.2 $<$ 0.5 verifying single emitter nature.}
\label{fig:SI_4}
\end{figure*}

The NV$^-$ excited state structure is sensitive to the local strain or electric field perturbations\cite{tamarat_spin-flip_2008, acosta_dynamic_2012, batalov_low_2009}. Once the strain perturbation exceeds the spin-orbit splittings, additional transverse strain increases the energy splitting between the Ex and Ey orbital branches whereas a perturbation parallel to the NV axis shifts the entire excited state manifold. We can observe the splitting under off-resonant excitation (532\,nm) (Fig.~\ref{fig:SI_4}a) in low temperature (T = 8\,K) spectra. Both device-coupled and bare-diamond NV centers exhibit varying magnitudes of splitting, from $<$\,8.2\,GHz (resolution limited by spectrometer) to $\sim$90\,GHz. The two transitions (E$_x$, E$_y$) exhibit near-orthogonal excitation polarization dependence (Fig.~\ref{fig:SI_4}a, inset). We confirm that the observed (E$_x$, E$_y$) pairs correspond to single NV centers by performing autocorrelation measurements (Fig.~\ref{fig:SI_4}b).\\

Finite element analysis (COMSOL Multiphysics) was used to verify whether the observed mean energy shift between device-coupled NVs and nearby bare-diamond NVs as well as the inhomogeneous broadening (main-text Fig. 3b) can be attributed to strain induced by the GaP membranes at \SI{8}{\kelvin}. To translate the components of the stress tensor to energy shift for each of the 4 NV orientations, we used the expressions given in Table~\ref{table:stress_shift}, where parameters $A_1$ and $A_2$ (following a negative sign convention for compressive stress/strain) are, respectively, $-1.47$, and
$\SI{3.85}{\milli\electronvolt\per\giga\pascal}$ \cite{davis_vibrionic_band_1976, smith_strain_2013}; for completeness, expressions for calculating the splitting between the Ex and Ey orbital branches are also provided in Table~\ref{table:stress_splitting}.

\begin{table}[h!]
\vspace{2em}
  \centering
  \begin{tabular}{c|l}
    \textbf{NV axis} & \textbf{Shift} \\
    \hline
    $[111]$ & $A_1(\sigma_{xx} + \sigma_{yy} + \sigma_{zz}) + 2A_2(\sigma_{xy} + \sigma_{yz} + \sigma_{zx})$ \\
    $[\bar 1\bar 11]$ & $A_1(\sigma_{xx} + \sigma_{yy} + \sigma_{zz}) + 2A_2(\sigma_{xy} - \sigma_{yz} - \sigma_{zx})$ \\
    $[\bar 1 1\bar 1]$ & $A_1(\sigma_{xx} + \sigma_{yy} + \sigma_{zz}) - 2A_2(\sigma_{xy} + \sigma_{yz} - \sigma_{zx})$ \\
    $[1\bar 1\bar 1]$ & $A_1(\sigma_{xx} + \sigma_{yy} + \sigma_{zz}) - 2A_2(\sigma_{xy} - \sigma_{yz} + \sigma_{zx})$ \\
  \end{tabular}
  \caption{Energy shift in terms of the components of the stress tensor in the crystal coordinates for different NV orientations.}
  \label{table:stress_shift}
\end{table}

\begin{table}[h!]
\vspace{1em}
  \centering
  \begin{tabular}{c|l}
    \textbf{NV axis} & \textbf{Splitting} \\
    \hline
    $[111]$ & $2\sqrt{3(B(\sigma_{xx}-\sigma_{yy})+C(\sigma_{yz}-\sigma_{zx}))^2+(B(\sigma_{xx}+\sigma_{yy}-2\sigma_{zz}) - C(2\sigma_{xy}-\sigma_{yz}-\sigma_{zx}))^2}$ \\
    $[\bar 1\bar 11]$ & $2\sqrt{3(B(\sigma_{xx}-\sigma_{yy})-C(\sigma_{yz}-\sigma_{zx}))^2+(B(\sigma_{xx}+\sigma_{yy}-2\sigma_{zz}) - C(2\sigma_{xy}+\sigma_{yz}+\sigma_{zx}))^2}$ \\
    $[\bar 1 1\bar 1]$ & $2\sqrt{3(B(\sigma_{xx}-\sigma_{yy})-C(\sigma_{yz}+\sigma_{zx}))^2+(B(\sigma_{xx}+\sigma_{yy}-2\sigma_{zz}) + C(2\sigma_{xy}-\sigma_{yz}+\sigma_{zx}))^2}$ \\
    $[1\bar 1\bar 1]$ & $2\sqrt{3(B(\sigma_{xx}-\sigma_{yy})+C(\sigma_{yz}+\sigma_{zx}))^2+(B(\sigma_{xx}+\sigma_{yy}-2\sigma_{zz}) + C(2\sigma_{xy}+\sigma_{yz}-\sigma_{zx}))^2}$ \\
  \end{tabular}
  \caption{Splitting between the Ex and Ey orbital branches in terms of the components of the stress tensor in the crystal coordinates for different NV orientations. $B =
    \SI{1.04}{\milli\electronvolt\per\giga\pascal}$,
    $C=\SI{1.69}{\milli\electronvolt\per\giga\pascal}$
    \cite{davis_vibrionic_band_1976, smith_strain_2013}}.
  \label{table:stress_splitting}
\end{table}
\hfill

\begin{figure*}[ht]
\centering
\includegraphics[width=5.2in]{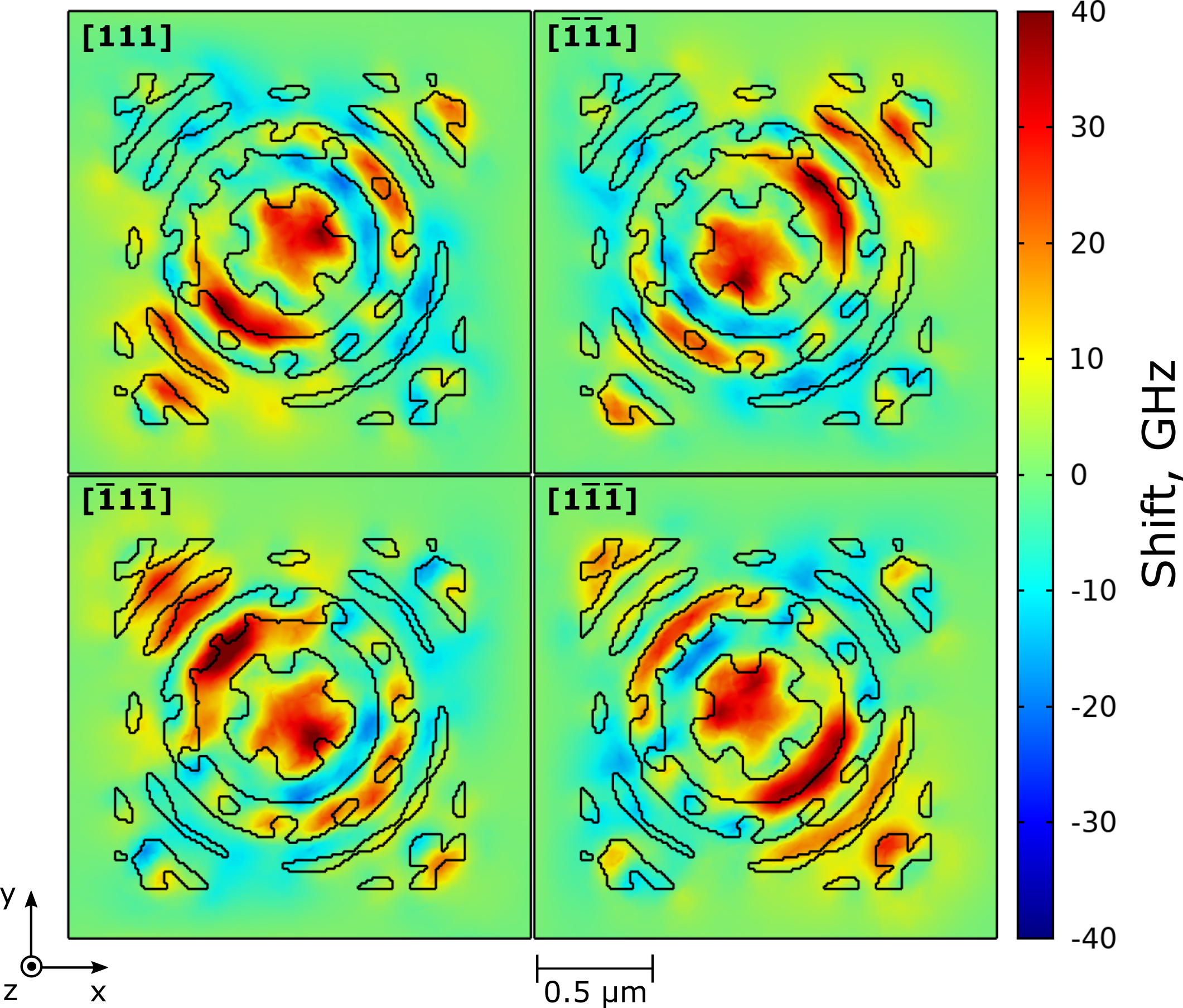}
\caption{Mean energy shift in GHz for different NV orientations at
  $\SI{100}{\nano\meter}$ from the surface underneath the GaP
  membrane. Finite element simulation is performed using COMSOL
  Multiphysics.}
\label{fig:shift100}
\end{figure*}

\begin{figure*}[ht]
\centering
\includegraphics[width=5.8in]{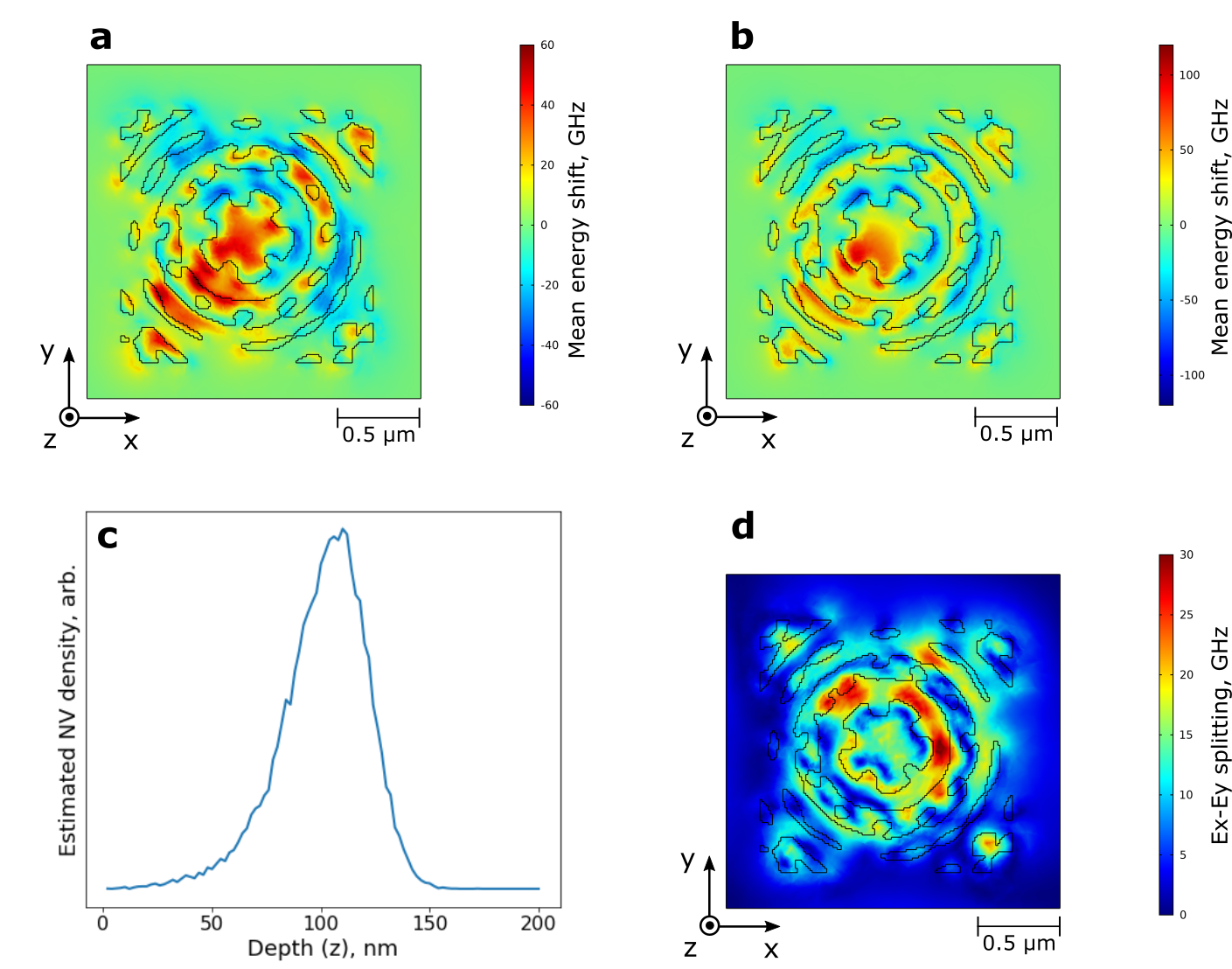}
\caption{\textbf{a.} Mean energy shift at $\SI{50}{\nano\meter}$ from the surface. \textbf{b.} Mean energy shift at $\SI{25}{\nano\meter}$ from the surface. \textbf{c.} Estimated NV density derived from the N ion distribution calculated according to the Kinchin-Pease model. \textbf{d.} Splitting between Ex and Ey orbital branches at $\SI{100}{\nano\meter}$ from the surface; the splitting due to strain from the Gap-diamond interface is comparable to that induced by the intrinsic strain from implantation which is consistent with our observation of both device-coupled and bare-diamond NV spectra. (\textbf{a}, \textbf{b} and \textbf{d} show the $[111]$ orientation; also note the differences in the scale of the color legends.)}
\label{fig:nv1_depths}
\end{figure*}

The dependence of the energy shift on the horizontal position at
$\SI{100}{\nano\meter}$ from the (100) surface (where the implanted NVs are most likely to be found) for all possible NV orientations is shown in Fig~\ref{fig:shift100}. The range of the color legend was chosen to be symmetric about $\SI{0}{\giga\hertz}$ for better visual reference. The actual shifts, however, range approximately from $\SI{-20}{\giga\hertz}$ to $\SI{40}{\giga\hertz}$. This, together with the fact that one of the most prominent positive shift regions occupies more than half of the area underneath the central feature of the photon extractors coupled to NVs of all 4 orientations, suggests that the average blue shift found in the ZPL wavelength distribution of device-coupled NVs relative to that of bare-diamond NVs and estimated at $\SI{30\pm15}{\giga\hertz}$ may be attributed to GaP-membrane-induced stress. The overall inhomogeneous broadening may be related to the NV distribution with respect to the distance from the surface (see Fig.~\ref{fig:nv1_depths}c): while most NVs are located at the depth of $\SI{100}{\nano\meter}$, the distribution's left tail extends almost to the surface (i.e., the depth of $\SI{0}{\nano\meter}$). Devices coupled to NVs about $\SI{30}{\nano\meter}$ from the surface can create stresses consistent with shifts ranging from $-80$ to $\SI{90}{\giga\hertz}$; for NVs even closer to the surface, this range gets significantly wider, as shown in Fig.~\ref{fig:nv1_depths} (a-b). It is also evident from the total relative size of the positive and negative shift regions that at depths $\leq\SI{100}{\nano\meter}$ NVs are somewhat more likely to exhibit a positive energy shift (assuming their uniform lateral distribution underneath the GaP membrane). For that reason, the distribution of ZPL wavelengths for device-coupled NVs would be expected to have a negative skewness (as is the case in Fig. 3b).

\vspace{3em}

The observed implanted NV (E$_x$, E$_y$) splitting (see Fig.~\ref{fig:nv1_depths}d for the estimated splitting at 100\,nm due to strain from the GaP-diamond interface) does not preclude their use for identical single-photon generation as it is within the capabilities demonstrated by Stark tuning \cite{acosta_dynamic_2012,schmidgall_frequency_2018}. However, the high strain regime can be detrimental to quantum information schemes that rely on high fidelity single-shot qubit spin readout~\cite{humphreys_deterministic_2018}. In a high transverse strain environment, the probability of a spin-flip during an optical transition increases, placing the NV$^-$ center in an effective dark state. This is illustrated in our resonant excitation measurements. To perform resonant excitation studies, we have to effectively mix the ground state $m_s=0$ and $m_s =\pm 1$ populations. We achieve this by adding 2.88\,GHz sidebands to our resonant scanning laser to simultaneously drive both spin states. 

\newpage
\section{Oxygen annealing devices post fabrication}

\begin{figure*}[h]
\centering
\includegraphics[width=5.5in]{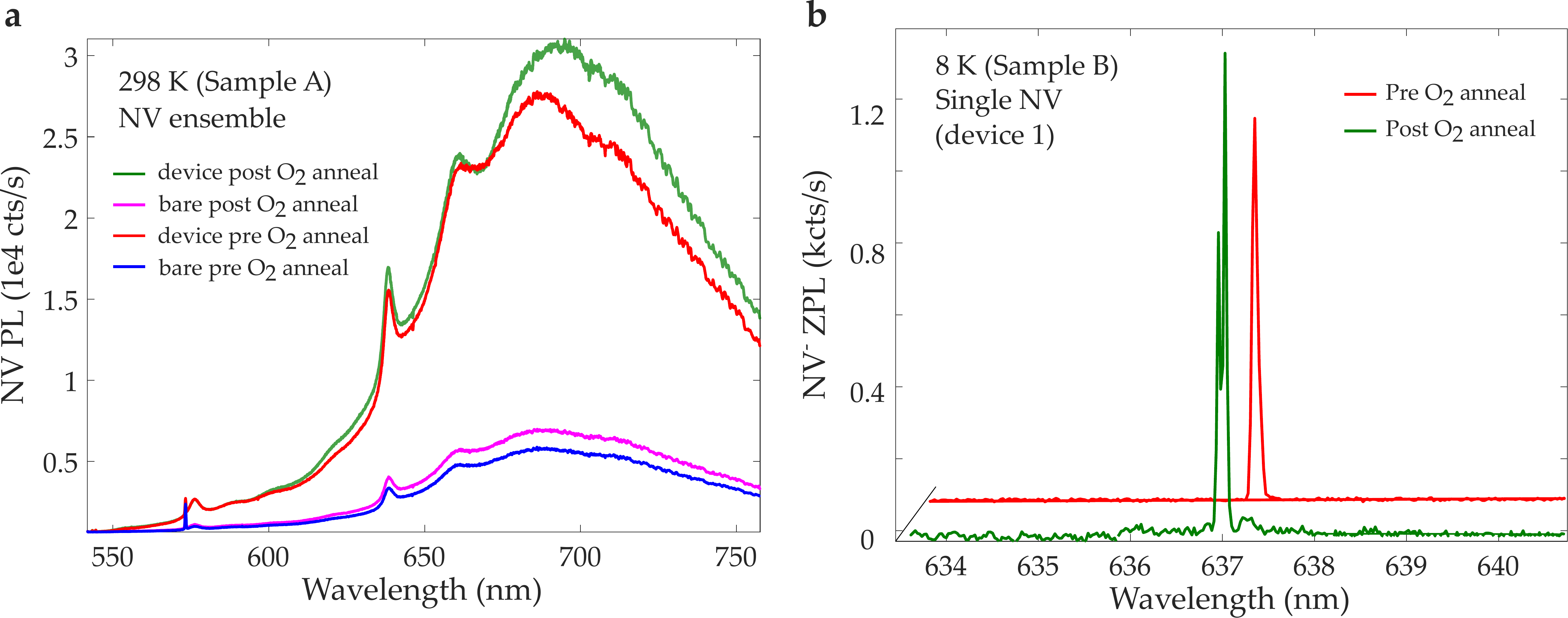}
\caption{\textbf{a.} Sample A NV PL spectra before and after oxygen annealing. \textbf{b.} Sample B device-coupled NV ZPL (device 1, shown in Fig.2 main text and SI.5) before and after oxygen annealing.}
\label{fig:oxygen_annealing}
\end{figure*}

Post fabrication and after measurements presented in Fig.2, both samples A and B were annealed at 400\,$^\circ$C for 4 hours under constant oxygen flow. On sample A, we observe an increase in overall NV PL emission (Fig.~\ref{fig:oxygen_annealing}a). The ensemble coupled devices show a similar trend. After oxygen annealing sample B, we do not observe any notable changes to PL intensity. The observed ZPL intensity for a device-coupled single NV (Fig.~\ref{fig:oxygen_annealing}b) is similar to pre-anneal. As discussed in the main text, we see a significant improvement in NV$^-$/NV$_0$ ratio, increased charge state stability and reduced spectral diffusion. Further annealing of sample A at 425\,$^\circ$C caused irreversible damage to the GaP layer destroying the photon extractors.

\bibliography{sri_references.bib, christians_refs.bib, andrews_refs.bib, pengning_refs.bib}   